\pdfoutput=1

\documentclass[sigconf]{acmart}

\acmConference[ICSE 2022]{The 44th International Conference on Software Engineering}{May 21–29, 2022}{Pittsburgh, PA, USA}
\AtBeginDocument{%
  \providecommand\BibTeX{{%
    \normalfont B\kern-0.5em{\scshape i\kern-0.25em b}\kern-0.8em\TeX}}}

\usepackage{framed}
\usepackage{listings}
\usepackage{xcolor}
\usepackage{color}
\usepackage[]{collab}
\usepackage{multirow}
\usepackage{algorithm}
\usepackage{algorithmic}
\usepackage{amsmath}
\usepackage{amssymb}
\usepackage{threeparttable}
\usepackage{url}
\usepackage[most]{tcolorbox}
\usepackage{hyperref}
\usepackage[hyphenbreaks]{breakurl}
\hypersetup{
    colorlinks = true,
    linkcolor=blue,
    filecolor=blue,      
    urlcolor=blue,
    citecolor=cyan,
}

\usepackage[shortlabels]{enumitem}
\setlist[enumerate]{nosep}

\definecolor{codegreen}{rgb}{0,0.6,0}
\definecolor{codegray}{rgb}{0.5,0.5,0.5}
\definecolor{codepurple}{rgb}{0.58,0,0.82}
\definecolor{backcolour}{rgb}{0.95,0.95,0.92}

\lstdefinestyle{mystyle}{
    commentstyle=\color{brown},
    keywordstyle=\color{magenta},
    numberstyle=\tiny\color{codegray},
    stringstyle=\color{codepurple},
    basicstyle=\ttfamily\footnotesize,
    breakatwhitespace=false,         
    breaklines=true,                 
    captionpos=b,                    
    keepspaces=true,                 
    numbers=left,                    
    numbersep=5pt,                  
    showspaces=false,                
    showstringspaces=false,
    showtabs=false,                  
    tabsize=2
}

\newcommand{\pytype}{Pytype\xspace}
\newcommand{\pysonar}{Pysonar2\xspace}
\newcommand{\pyre}{Pyre Infer\xspace}
\newcommand{\typilus}{Typilus\xspace}

\newcommand{\typepy}{Type4Py\xspace}
\newcommand{\tool}{\textsc{HiTyper}\xspace}
\newcommand\etal{{\it{et al.\ }}}

\newcommand{\tabincell}[2]{\begin{tabular}{@{}#1@{}}#2\end{tabular}}
\newcommand{\eat}[1]{\if 0 #1 \fi}

\copyrightyear{2022}
\acmYear{2022}
\setcopyright{acmcopyright}\acmConference[ICSE '22]{44th International Conference on Software Engineering}{May 21--29, 2022}{Pittsburgh, PA, USA} \acmBooktitle{44th International Conference on Software Engineering (ICSE '22), May 21--29, 2022, Pittsburgh, P A, USA}
\acmPrice{15.00}
\acmDOI{10.1145/3510003.3510038}
\acmISBN{978-1-4503-9221-1/22/05}

\newfont{\mycrnotice}{ptmr8t at 7pt}
\newfont{\myconfname}{ptmri8t at 7pt}
\begin{document}

%%
%% The "title" command has an optional parameter,
%% allowing the author to define a "short title" to be used in page headers.
\title{Static Inference Meets Deep Learning: A Hybrid Type Inference Approach for Python}
% Improving Neural Type Prediction with Static Type Inference
% \tool: A Hybrid Static Type Inference Framework \\ with Neural Prediction

\author{Yun Peng}
 \affiliation{\institution{The Chinese University of Hong Kong}\country{Hong Kong, China}}
 \email{ypeng@cse.cuhk.edu.hk}
 
 \author{Cuiyun Gao}
 \affiliation{\institution{Harbin Institute of Technology} \country{Shenzhen, China}}
 \email{gaocuiyun@hit.edu.cn}
 \authornote{Corresponding author}

 \author{Zongjie Li}
 \affiliation{\institution{Harbin Institute of Technology }\country{Shenzhen, China}}
 \email{lizongjie@stu.hit.edu.cn}
 
 \author{Bowei Gao}
 \affiliation{\institution{Harbin Institute of Technology }\country{Shenzhen, China}}
 \email{1160300103@hit.edu.cn}

 \author{David Lo}
 \affiliation{\institution{Singapore Management University} \country{Singapore}}
 \email{davidlo@smu.edu.sg}
 
 \author{Qirun Zhang}
 \affiliation{\institution{Georgia Institute of Technology} \country{United States}}
 \email{qrzhang@gatech.edu}

 \author{Michael Lyu}
 \affiliation{\institution{The Chinese University of Hong Kong} \country{Hong Kong, China}}
 \email{lyu@cse.cuhk.edu.hk}
%%
%% The "author" command and its associated commands are used to define
%% the authors and their affiliations.
%% Of note is the shared affiliation of the first two authors, and the
%% "authornote" and "authornotemark" commands
%% used to denote shared contribution to the research.

%%
%% By default, the full list of authors will be used in the page
%% headers. Often, this list is too long, and will overlap
%% other information printed in the page headers. This command allows
%% the author to define a more concise list
%% of authors' names for this purpose.

%%
%% The abstract is a short summary of the work to be presented in the
%% article.
\begin{abstract}
Type inference for dynamic programming languages such as Python is an important yet challenging task. Static type inference techniques can precisely infer variables with enough static constraints but are unable to handle variables with dynamic features. Deep learning (DL)  based approaches are feature-agnostic,  
but they cannot guarantee the correctness of the predicted types. Their performance significantly depends on the quality of the training data (i.e., DL models perform poorly on some common types that rarely appear in the training dataset). It is interesting to note that the static and DL-based approaches offer complementary benefits. Unfortunately, to our knowledge, precise type inference based on both static inference and neural predictions has not been exploited and remains an open challenge. In particular, it is hard to integrate DL models into the framework of rule-based static approaches.

This paper fills the gap and  proposes a hybrid type inference approach named \tool based on both static inference and deep learning. Specifically, our key insight is to record type dependencies among variables in each function  and encode the dependency information in \emph{type dependency graphs} (TDGs). Based on TDGs, we can easily integrate type inference rules in the nodes to conduct static inference and type rejection rules to inspect the correctness of neural predictions. \tool iteratively conducts static inference and DL-based prediction until the TDG is fully inferred. Experiments on two benchmark datasets show that \tool outperforms state-of-the-art DL models by exactly matching 10\% more human annotations. \tool also achieves an increase of more than 30\% on inferring rare types. Considering only the static part of \tool, 
it infers 2$\times$ $\sim$ 3$\times$ more types than existing static type inference tools. Moreover,  \tool successfully corrected seven wrong human annotations in six GitHub projects, and two of them have already been approved by the repository owners.

\end{abstract} 

\maketitle

\section{Introduction}\label{sec:intro}
Dynamically typed programming languages such as Python are becoming increasingly prevalent in recent years. According to GitHub Octoverse 2019 and 2020~\cite{octoverse}, 
Python outranks Java and C/C++ and becomes one of the most popular programming languages. The dynamic features provide more flexible coding styles and enable fast prototyping. However, without concretely defined variable types, dynamically typed programming languages face challenges in ensuring security and compilation performance. According to a recent survey by Jetbrains~\cite{jetbrainssurvey}, static typing or at least some strict type hints becomes the top 1 desired feature among Python developers. To address such problems, 
some research adopts design principles of statically typed programming languages~\cite{Ray17study,Hanenberg13study,typesurvey}. For example, reusing compiler backend of the statically typed languages~\cite{numba} and predicting types for most variables~\cite{javainfer05,deeptyper,maxsmtpython18,rubyinfer09,jsinfer09,typewriter,typilus}. Moreover, Python  officially supports type annotations in the Python Enhancement Proposals (PEP)~\cite{pep484,pep544,pep585,pep589}.

Type prediction is a popular task performed by existing work. Traditional \textit{static type inference} approaches~\cite{javainfer05,maxsmtpython18,rubyinfer09,jsinfer09,jsnice} and type inference tools such as \pytype~\cite{pytype}, \pysonar~\cite{pysonar}, and \pyre~\cite{pyre} can correctly infer
types for the variables with enough static constraints, e.g., for \texttt{a = 1} we can know the type of \texttt{a} is \texttt{int}, but are unable to handle the variables with few static constraints, e.g. most function arguments or dynamic evaluations such as \texttt{eval()}~\cite{jsdynamic}. %\qirun{I think you should cite a PLDI'10 paper: An analysis of the dynamic behavior of JavaScript programs. see Gregor Richards' dblp page. I think there should also be a followup ECOOP paper. }

% \textit{Dynamic type inference} techniques~\cite{rubydinfer11,rubychecker13} simulate the workflow of functions and predict types according to input cases. Although these techniques generally perform better than static type inference, they suffer from limited code coverage and huge time consumption. Thus, dynamic type inference techniques are difficult to be deployed on large code bases. \qirun{You need to be more careful here. For Javascript, V8 based on feedback-based type specialization is very fast. There should also be a few PLDI papers on this topic. I didn't precisely recall. You should definitely discuss/survey javascript.  }

With the recent development of \textit{deep learning (DL) methods}, we can leverage more type hints such as identifiers and existing type annotations to predict types. Many DL-based methods~\cite{pbinfer16, deeptyper, typewriter, typilus,NL2Type, mir2021type4py} have been proposed, and they show significant improvement compared with static techniques~\cite{Le20survey}. While DL-based methods are effective, they face the following two major limitations:
\begin{enumerate}[leftmargin=*,wide, label=(\roman*)]
\item \emph{No guarantee of the type correctness}. Pradel \etal~\cite{typewriter} find that the predictions given by DL models are inherently imprecise as they return a list of type candidates for each variable, among which only one type is correct under a certain context. Besides, the predictions made by DL models may contradict the typing rules, leading to type errors. Even the state-of-the-art DL model Typilus~\cite{typilus} generates about 10\% of predictions that cannot pass the test of a type checker. The type correctness issue makes the DL-based methods hard to be directly deployed into large codebases without validation. Recent work~\cite{typewriter, typilus} leverages a search-based validation in which a type checker is used to validate all combinations of types returned by DL models and remove those combinations containing wrong types. However, these approaches cannot correct the wrong types but only filter them out.

\item \emph{Inaccurate prediction of rare types}. Rare types refer to the types with low occurrence frequencies in datasets~\cite{typilus}. Low-frequency problem has become one of the bottlenecks of DL-based methods~\cite{zhang-etal-2019-long,raunak-etal-2020-long,liu2020deep,ren2020balanced,kang2020decoupling}. For example, Typilus's accuracy drops by more than 50\% for the types with occurrence frequencies fewer than 100, compared to the accuracy of the types with occurrence frequencies more than 10,000. More importantly, rare types totally account for a significant amount of annotations even though each of them rarely appears. We analyze the type frequencies of two benchmark datasets from Typilus~\cite{typilus} and Type4Py~\cite{type4pydataset}, and find a \emph{long tail phenomenon}, i.e., the top 10 types in the two datasets already account for 54.8\% and 67.8\% of the total annotations, and more than 10,000 and 40,000 types in two datasets are rare types with frequency proportions less than 0.1\%. They still occupy 35.5\% and 25.5\% of total annotations for the two respective datasets and become the long ``tail'' of type distributions.
\end{enumerate}
% The top 10 types in the two datasets account for 54.8\% and 67.8\% of total annotations, but there are 12,138 types and 42,290 types in the two datasets, respectively. If we adopt the frequency threshold of Typilus (0.1\%) to identify rare types, most types are rare types and they totally occupy 35.5\% and 25.5\% for the two datasets. 
%Given the significant performance drop of DL-based methods on rare types and substantial amount of rare types in real-world programs, accurately predicting the rate types becomes one essential problem for the task.

% as a passive search-based method, this approach still inherits the rare type prediction problem as it can never correct wrong types but only filter out them. This integration does not fully utilize the inference ability of static typing rules.

% Noting that static type inference and DL models can work together to address problems of each other
To remedy the limitations of the previous studies, this paper  proposes a hybrid type inference framework named \tool, which conducts static type inference and accepts recommendations from DL models (\textit{Static+DL}). We propose a novel representation, named type dependency graph (TDG), for each function, where TDG records the type dependencies among variables. Based on TDG, we reformulate the type inference task into a blank filling problem where the ``blanks'' (variables) are connected with dependencies so that both static approaches and DL models can fill the types into ``blanks''. 

\tool infers the ``blanks'' in TDG mainly based on static type inference, which automatically addresses DL models' rare type prediction problem since static type inference rules are insensitive to type occurrence frequencies. \tool extends the inference ability of static type inference by accepting recommendations from DL models when it encounters some ``blanks'' that cannot be statically inferred. Different from the search-based validation by Pradel \etal~\cite{typewriter}, \tool builds a series of type rejection rules to filter out all wrong predictions on TDG, and then continues to conduct static type inference based on the reserved correct predictions.

We evaluate \tool on two public datasets. One dataset is released by Allamanis \etal in the paper of Typilus~\cite{typilus}, and the other is ManyTypes4Py~\cite{mir2021type4py}, one large dataset recently
released for this task. Experiment results show that \tool outperforms both SOTA DL models and static type inference tools. Compared with two SOTA DL models Typilus and Type4Py, \tool presents a 10\%$\sim$12\% boost on the performance of overall type inference, and a 6\% $\sim$ 71\% boost on the performance of certain kinds of type inference such as return value type inference and user-defined type inference. Without the recommendations from neural networks and only looking at the static type inference part, \tool generally outputs 2$\times$ $\sim$ 3$\times$ more annotations with higher precision than current static type inference tools Pyre~\cite{pyre} and Pytype~\cite{pytype}. \tool can also identify wrong human annotations in real-world projects. We identify seven wrong annotations in six projects of \typilus's dataset and submit pull requests to correct these annotations. Two project owners have approved our corrections. 

\textbf{Contributions.} Our contributions can be concluded as follows:
\begin{itemize}
    \item To the best of our knowledge, we are the first to propose a hybrid type inference framework that integrates static inference with DL for more accurate type prediction.
    \item We design an innovative type dependency graph to strictly maintain type dependencies of different variables.
    \item We tackle some challenges faced by previous studies and design a series of type rejection rules and a type correction algorithm to validate neural predictions.
    \item Extensive experiments demonstrate the superior performance of the proposed \tool than SOTA baseline models and static type inference tools in the task.
\end{itemize}

\begin{figure}[t]
    \setlength{\belowcaptionskip}{-0.5cm}
    \centering
    \includegraphics[width = 0.4 \textwidth]{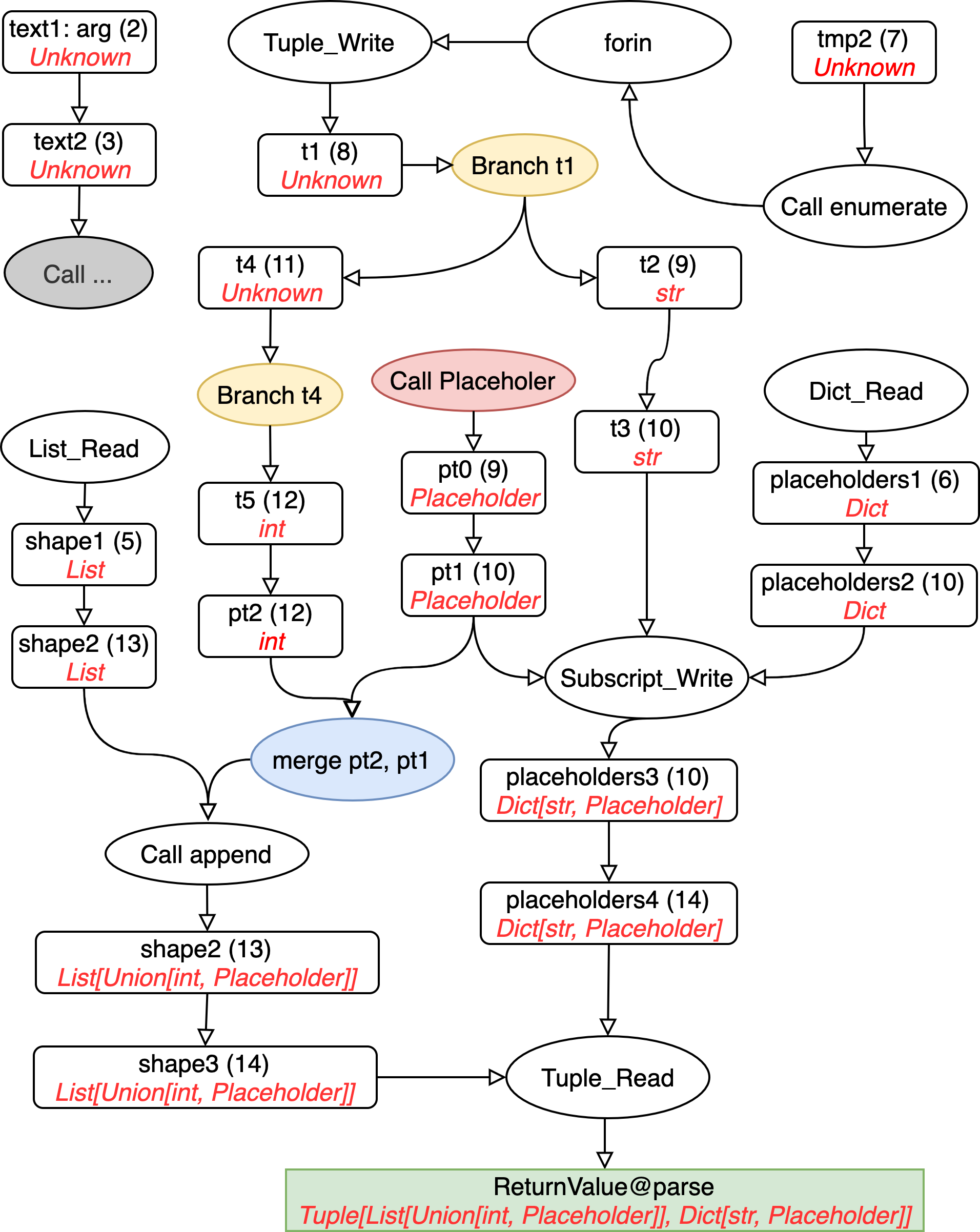}
    \caption{Type dependency graph of the parse() from Code.\ref{lst:mot_ex}.}
    \label{fig:mot_typegraph}
\end{figure}

\section{Motivating Example}\label{sec:motivation}

\eat{
\textbf{Importance of Rare Type Prediction.} To capture the general distribution of types and avoid potential bias, we analyze two public datasets. The first one is Typilus's dataset released by Allamanis \etal~\cite{typilus}, and it contains 153,731 annotations with 12,138 different types from 592 Python projects. The second dataset is ManyTypes4Py released by A. M. Mir \etal~\cite{mir2021type4py}. ManyTypes4Py is much larger than Typilus's dataset, and includes 758,129 annotations with 42,290 different types from 5,203 Python projects. We show the accumulative distribution of annotations with respect to the proportion of types in Fig.~\ref{fig:typedist}.

From Fig.~\ref{fig:typedist}, we can observe that both datasets present long-tail distribution, i.e., the first 95\% of types only accounts for about 20\% of annotations.  What's more, The top 10 types account for 54.8\% and 67.8\% of all the annotations for Typilus's dataset and ManyTypes4Py, respectively. This indicate that only a very small proportion of types accounts for a very large proportion of annotations and a large amount of types have quite low frequency in the dataset. However, even if many types have quite low frequency in the dataset, the total proportion of annotations is not small. If we adopt the frequency threshold of Typilus (0.1\%) to identify rare types, they occupy  i.e., 35.5\% and 25.5\% for the two datasets, respectively. DL models are proven ineffective to predict the rare types. For example, \typilus only accurately predicts $<$30\% of the rare types, but $>$80\% of common types are successfully predicted~\cite{typilus}. Therefore, Rare types become a bottleneck if we want to further improve the performance on this task. Improving the prediction performance of rare types is essential for the type prediction task.}

% share quite similar type distribution, even though the second dataset is much larger than the first one. What's more important, the number of the type annotations drops dramatically from the first most popular type \texttt{str} to the tenth most popular type \texttt{Optional[str]}. Some basic built-in types like \texttt{str}, \texttt{int}, \texttt{bool}, and \texttt{float} are quite commonly used and account for about 55\% $\sim$ 60\% of annotations. However, the remaining thousands of types are rarely used since each of them accounts for less than 0.5\% of the total annotations. This indicates a \textit{Long Tail phenomenon}: quite a few basic built-in types are commonly used but a substantial number of other types, which totally account for $\sim$40\% of the total annotations, have low occurrence frequency. When maually checking these rare types, we find that most of these types are user-defined types and composite types, which only occur in a few source files and can vary across projects, making it hard to include them in a fixed type set and for DL models to learn the features. For example, \typilus successfully predicts types of over 80\% common types, but only successfully predicts types of less than 30\% rare types~\cite{typilus}. As rare types actually account for a substantial amount of annotations, it has become the major barrier that hinders the performance of DL models to be further improved. 

Listing~\ref{lst:mot_ex} illustrates an example of code snippet from the WebDNN project.\footnote{\url{https://github.com/mil-tokyo/webdnn}} Results of several baselines, including static type inference techniques - Pytype and Pysonar2, and state-of-the-art DL models - Typilus, are depicted in Table~\ref{tab:mot_ex}.

\begin{lstlisting}[language = python,caption = A Function from WebDNN. , label = lst:mot_ex]
#src/graph_transpiler/webdnn/graph/shape.py
def parse(text):
    normalized_text = _normalize_text(text)
    tmp = ast.literal_eval(normalized_text)
    shape = []
    placeholders = {}
    for i, t in enumerate(tmp):
        if isinstance(t, str):
            pt = Placeholder(label=t)
            placeholders[t] = pt
        elif isinstance(t, int):
            pt = t
        shape.append(pt)
    return shape, placeholders
\end{lstlisting}

\textbf{Static Inference.}
According to Table~\ref{tab:mot_ex}, we can find that the static type inference techniques fail to infer the type of the argument \texttt{text} since the argument is at the beginning of data flow without any assignments or definitions. One common solution to infer the type is to use inter-procedural analysis and capture the functions that call \texttt{parse()}~\cite{pycg}. However, tracing the functions in programs, especially in some libraries, is not always feasible. As for the return value, by analyzing the data flow and dependencies between variables, static inference can easily identify that \texttt{shape} (line 5, 13) and \texttt{placeholders} (line 6, 10) consist of the return value. It can recursively analyze the types of the two variables, and finally output the accurate type of the return value. Indeed,  both Pysonar2 and Pytype can correctly infer that the return value is a tuple containing a list and dict.

\textbf{DL Approach.}
The DL model Typilus~\cite{typilus} accurately predicts the type as \texttt{str} according to the semantics delivered by the argument \texttt{text} and contextual information. The case illustrates that DL models can predict more types than static inference. However, Typilus fails to infer the type of the return value of \texttt{parse()}. Current DL models cannot maintain strict type dependencies between variables. Therefore, Typilus only infers the type as a tuple but cannot accurately predict the types inside the tuple. When adding a type checker to validate Typilus's predictions, its argument prediction is reserved since it does not violate any existing type inference rules. However, for the return value, its 2nd and 3rd type predictions in Table~\ref{tab:mot_ex} by Typilus are rejected since the return value of \texttt{parse()} explicitly contains two elements with different types. The 1st prediction is also rejected because it contains the type \texttt{Optional[text]} that does not appear in the return value. In this case, the model does not produce any candidate type for the return value.

\textbf{Static+DL Approach.}
% Static inference + DL models.
For the code example, we find that static inference is superior than DL models when sufficient static constraints or dependencies are satisfied, while DL models are more applicable for the types lacking sufficient static constraints. Given the code, \tool first generates the TDG of it, as shown in Fig.~\ref{fig:mot_typegraph}, and tries to fill all nodes in TDG with corresponding types ("blank filling"). For the argument \texttt{text}, \tool identifies that the type cannot be inferred by static inference (it does not have any input edges) and asks DL for recommendations. \tool does not directly output the predictions from DL as final type assignments. Instead, \tool validates the prediction's correctness and accepts the result only if no type inference rules are violated. When predicting the return value, \tool captures its type dependencies based on the TDG (it connects with two input nodes) and directly leverages static inference to infer the type. For this case, DL predictions are not required, largely avoiding the imports of wrong types.

\begin{table}[t]
    \centering
    \caption{Prediction results of different baselines for Listing~\ref{lst:mot_ex}.}
    \scalebox{0.85}{\begin{tabular}{llcl}
    \toprule
    \textbf{Approach} & \textbf{Baseline} & \textbf{Argument} & \textbf{Return Value} \\
    \hline
    \hline
         & \multirow{2}*{\tabincell{l}{Ground \\ Truth}} & \multirow{2}*{\color{codegreen}{str}}  &  \multirow{2}*{\tabincell{l}{\color{codegreen}{Tuple[List[int, Placeholder],} \\ \color{codegreen}{Dict[str, Placeholder]]}}}\\
         & & \\
    \hline
         \multirow{2}{*}{\textbf{Static}} & \pysonar & ?  & Tuple[List[int],Dict]\\ 
         & \pytype & ? & Tuple[List, Dict] \\
    \hline
         \multirow{6}*{\textbf{DL}} & \multirow{6}*{\tabincell{c}{\typilus }} & \multirow{6}*{\color{codegreen}{1. str}}  & \multirow{4}*{\tabincell{l}{1. Tuple[collections.OrderedDict[ \\ Text, List[DFAState]], \\ Optional[Text]],Tuple[Any, \\ List[Tuple[Any]], Any]}} \\
         & & &  \\
         & & & \\
         & & & \\
         & & & 2. Tuple[Text] \\
         & & & 3. Tuple[torch.Tensor] \\
    %\hline
    %     \tabincell{l}{\textbf{DL} + \\ \textbf{Valid.}} & \tabincell{l}{Typilus \\ with valid.} & \color{codegreen}{str} & ? \\
    \hline
         \multirow{2}*{\tabincell{l}{\textbf{Static} \\ + \textbf{DL}}}& \multirow{2}*{\tabincell{l}{\tool\\ (\typilus)}} & \multirow{2}*{\color{codegreen}{str}}  &  \multirow{2}*{\tabincell{l}{\color{codegreen}{Tuple[List[int, Placeholder],} \\ \color{codegreen}{Dict[str, Placeholder]]}}} \\
         & & \\
    \bottomrule
    \end{tabular}
    }
    %\vspace{-1.5em}
    \label{tab:mot_ex}
\end{table}

\section{\tool}\label{sec:approach}
In this section, we first introduce the definitions used in \tool and then elaborate the details of \tool.

\begin{figure*}[htb]
    \setlength{\belowcaptionskip}{-0.5cm}
    \centering
    \includegraphics[width = 1.0 \textwidth]{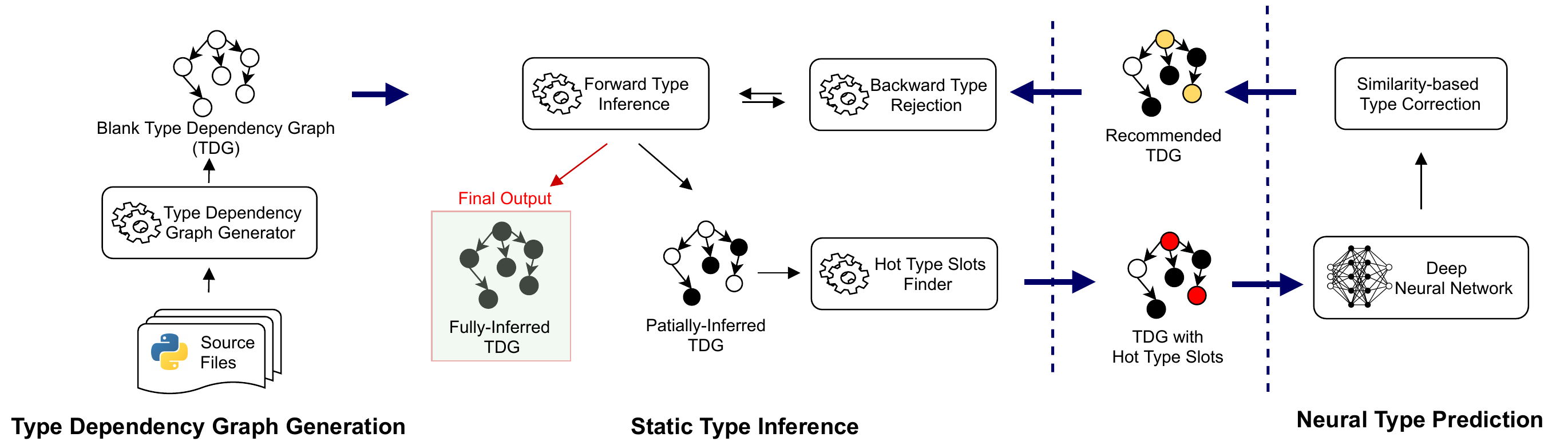}
    \caption{Overall architecture of \tool. Black solid nodes,  hollow nodes, red nodes and yellow nodes in the type dependency graphs represent inferred type slots, blank type slots, hot type slots, and the type slots recommended by DL model, respectively.}
    \label{fig:arch}
\end{figure*}

\subsection{Definition of Types}
Fig.~\ref{fig:type} shows the definitions of different types according to the official documentation of Python~\cite{py3doc} and its type checker \texttt{mypy}~\cite{mypydoc}. Note that we remove the \texttt{object} type and \texttt{Any} type since they are not strict static types. In general, all types can be classified into built-in types and user-defined types. Built-in types are predefined in the language specification of Python while user-defined types are created by developers. Developers can define the operations or methods supported by a user-defined type and overwrite some built-in operations for their user-defined types. For example, developers can define an \texttt{\_\_add\_\_()} method in a class so that two types derived from this class can be directly added together using the built-in operator \texttt{+}. The operation is called \textit{operator overloading}. We create a subcategory for user-defined types with operator overloading behaviors since they have different type inference rules.

The type categories showed in Fig.~\ref{fig:type} are widely used in most static type inference techniques~\cite{mypy, pyre, pytype}. Differently, DL-based studies~\cite{typilus,mir2021type4py} generally categorize the types into \textit{common types} and \textit{rare types} based on a pre-defined threshold of occurrence frequencies (e.g., 100 in~\cite{typilus}). For a fair comparison, we also follow this definition for evaluation. By analyzing the rare types in two public datasets \typilus and ManyTypes4Py, we find that 79.02\% and 99.7\% of rare types actually are user-defined types. Because static inference technique is frequency-insensitive and cannot recognize rare types, we mainly add supports for user-defined types on static inference side of \tool.

\begin{figure}[ht]
    \setlength{\belowcaptionskip}{-0.3cm}
    \setlength{\belowcaptionskip}{-0.6cm}
    \centering
    \begin{minipage}[t]{0.4\textwidth}
    \begin{align*}
    \theta \in \mathit{Type} \ (\Theta):: = & \  \gamma  \ | \ \alpha[\theta,...,\theta] \ | \ u \ | \ \textbf{None} \ | \ \textbf{type}  \\
    \gamma \in \mathit{Elementary} \ \mathit{Type} \ (\Gamma) ::= & \ \textbf{int} \ | \ \textbf{float} \ | \  \textbf{str} \ | \ \textbf{bool} \ | \ \textbf{bytes}   \\
    \alpha \in \mathit{Generic} \ \mathit{Type}  \ 
    (A)::= & \ \textbf{List} \ | \ \textbf{Tuple} \ | \ \textbf{Dict} \ | \ \textbf{Set} \ | \\ & \textbf{Callable} \ | \  \textbf{Generator} \ | \ \textbf{Union} \\
    b \in \mathit{Builtin} \ \mathit{Type} \ (B) ::= & \ \gamma \ | \ \alpha[\theta] \\
    u \in \mathit{User} \ \mathit{Defined}\ \mathit{Type} \ (U) & ::=  \mathit{all} \ \mathit{classes} \ \mathit{and} \ \mathit{named} \\ & \qquad \mathit{tuples} \ \mathit{in} \ \mathit{code} \\
    o \in \mathit{Overloading} \ \mathit{User} & ::= all \ classes \ with \\
     \ \mathit{Defined} \ \mathit{Type} \  (O) & \qquad  \mathit{operator} \ \mathit{overloading} \ \mathit{in} \ \mathit{code}
    \end{align*}
    \end{minipage}
\caption{Types in Python.}
\label{fig:type}

\end{figure}

\subsection{Overview}
\tool accepts Python source files as input and outputs JSON files recording the type assignment results. Fig.~\ref{fig:arch} illustrates its overall architecture. \tool includes three major components: type dependency graph generation, static type inference, and neural type prediction. The static type inference component comprises two main steps, i.e., forward type inference and backward type rejection. 

\textbf{Type Dependency Graph (TDG) Generation.}
Specifically, given a Python source file, \tool first generates TDGs for each function and identifies all the imported user-defined types (Sec.~\ref{sec:typegraph}). TDG transforms every variable occurrence and expression into nodes and maintains type dependencies between them so that static inference and DL models can work together to fill types into it.

\textbf{Static Type Inference - Forward Type Inference.}
To maintain the correctness of prediction results, \tool focuses on inferring types using static inference. Given a TDG, \tool conducts forward type inference by walking through the graph and implementing the type inference rules saved in each expression node (Sec.~\ref{sec:typeinfer}). However, due to the limitation of static inference, in most cases \tool can only infer partial type slots, i.e., variables, indicated as black solid nodes in the \textit{partially-inferred TDG} in Fig.~\ref{fig:arch}; while the blank nodes denote the type slots without sufficient static constraints and remaining unsolved. To strengthen the inference ability of \tool, we ask DL models for recommendations.

\textbf{Neural Type Recommendation.}
Through the \emph{hot type slot finder}, \tool identifies a key subset of the blank nodes as hot type slots, marked as red nodes in Fig.~\ref{fig:arch}, for obtaining recommendations from DL models. \tool also employs a similarity-based type correction algorithm to supplement the prediction of user-defined types, which are the primary source of rare types (Sec.~\ref{sec:typereco}). The types recommended by the neural type prediction component are filled into the graph, resulting in the \textit{recommended TDG}. 

\textbf{Static Type Inference - Backward Type Rejection.}
\tool utilizes type rejection rules to validate the neural predictions in hot type slots (Sec.~\ref{sec:typeinfer}). Then it traverses the whole TDG to transmit the rejected predictions from output nodes to input nodes so that all nodes in TDG can be validated. Finally \tool invokes forward type inference again to infer new types based on the validated recommendations.

The interactions between forward type inference and backward type rejection could iterate until the TDG reaches a fixed point, i.e., the types of all nodes do not change any more. Meanwhile, the iterations between static inference and neural prediction can repeat several times until all type slots are inferred, or a maximum iteration limit is reached.

% Some type slots are inferred at this step (black solid nodes in the type graphs in Fig.~\ref{fig:arch}); however, most of them without appropriate static constraints remain unsolved, i.e., blank type slots. \tool identifies a key subset of them as hot type slots (red nodes in Fig.\ref{fig:arch}) and then adopts a SOTA DL model to recommend candidate types for the hot type slots. To supplement user-defined types which are not in the DL model's training set, \tool also employs a similarity-based type correction algorithm (Sec.~\ref{sec:typereco}) to find the most similar valid user-defined type and replace the one predicted by the DL model. The recommended types are filled into the type dependency graph. \tool utilizes type rejection rules (Sec.~\ref{sec:typerej}) to validate the neural predictions and then conducts forward type inference again. The process between forward type inference and backward type rejection could iterate until the type dependency graph reaches a fixed point, i.e., types of all nodes do not change any more. Meanwhile, the whole iterations between static inference and neural prediction can repeat several times until all type slots are inferred.

% In the following subsections, we will elaborate the main components in \tool: type dependency graph, forward type inference, backward type rejection, and type recommendations. 

\subsection{Type Dependency Graph Generation}\label{sec:typegraph}

This section introduces the creation of type dependency graph (TDG), which describes the type dependencies between different variables in programs. Fig.~\ref{fig:expr} presents the syntax of all the expressions that generate types in Python, where each expression corresponds to a node in the AST (Abstract Syntax Tree). Given the AST of a program, \tool can quickly identify these expressions. The expression nodes constitute a major part of TDG. We define TDG as below.

% \tool utilizes type dependency graph to represent the type dependencies between different variables in source code. Before introducing the definition of type dependency graph, we first give a look at all expressions that generates types in Python. The syntax of these expressions are shown in Fig.~\ref{fig:expr}. There is a special AST node for each expression we show in Fig.~\ref{fig:expr}, so \tool can easily identify these expressions given the AST of input code. These expression nodes consist of the major part of our type dependency graph

\textbf{Definition.} 
We define a graph $G = (N, E)$ as a type dependency graph (TDG), where $N = \{n_{i}\}$ is a set of nodes representing all variables and expressions in source code, and $E$ is a set of directed edges of $n_i \rightarrow n_j$ indicating the type of $n_j$ can be solved based on the type of $n_i$ by type inference rules. We also denote $n_i$ is the input node of $n_j$ and $n_j$ is the output node of $n_i$ here.

\begin{figure}[th]
    \setlength{\belowcaptionskip}{-0.5cm}
    \centering
    \begin{minipage}[t]{0.4\textwidth}
    \begin{align*}
    e \in \mathit{Expr} & ::= v \ | \ c \ | \ e \ \textbf{blop} \ e \ |\  e \ \textbf{numop} \ e \ | \\ &  \ e \ \textbf{cmpop} \ e \ | \ e \ \textbf{bitop} \ e \ | \\  & \ (e,...,e) \ | \ [e,...,e] \ | \\ & \ \{e:e,...,e:e\} \ | \ \{e,...,e\} \ | \\ &  [e \ \textbf{for} \ e \ \textbf{in} \ e] \ | \ \{e \ \textbf{for}\ e \ \textbf{in} \ e\} \ | \\& \{e:e \ \textbf{for} \ e, e \ \textbf{in} \ e\} \ | \ (e \ \textbf{for} \ e \ \textbf{in} \ e) \ | \\ & e(e,...,e) \ | \ e[e:e:e] \  | \ e.v \\
    v \in \mathit{Variables} & ::= \mathit{all} \ \mathit{identifiers} \ \mathit{in} \ \mathit{code} \\
    c \in \mathit{Constants} & ::= \mathit{all} \ \mathit{literals} \ \mathit{in} \ \mathit{code} \\
    blop \in \mathit{Boolean} \ \mathit{Operations} & :: = And \ |\ Or \ | \ Not\\
    numop \in \mathit{Numeric} \ \mathit{Operations} & ::= Add \ |\ Sub \ | \ Mult \ | \ Div \ | \ Mod \ | \\ & \ UAdd \ | \ USub  \\
    bitop \in \mathit{Bitwise} \ \mathit{Operations} & ::= LShift \ | \ RShift \ | \ BitOr \ | \ BitAnd \ | \\ & \ BitXor \ | \ FloorDiv \ | \ Invert \\
    cmpop \in \mathit{Compare} \ \mathit{Operations} & ::= Eq \ | \ NotEq \ | \ Lt \ | \ LtE \ | \ Gt \ | \ GtE \ | \\ & \ Is \ | \ IsNot \ | \ In \ | \ NotIn
\end{align*}
\end{minipage}
\caption{The syntax of expressions for typing in Python}
\label{fig:expr}
\end{figure}

The TDG contains four kinds of nodes:
\begin{itemize}
    \item \textit{symbol} nodes represent all the variables for which the types need to be inferred. We also use \textit{type slots} to indicate symbol nodes in the following sections.
    \item \textit{expression} nodes represent all the expressions that generate types as shown
    in Fig.~\ref{fig:expr}.
    \item \textit{branch} nodes represent the branch of data flows.
    \item \textit{merge} nodes represent the merge of data flows.
\end{itemize}

\tool creates a node for every variable occurrence instead of every variable in TDG because Python's type system allows variables to change their types at run-time. Similar to static single assignment (SSA), \tool labels each occurrence of a variable with the order of occurrences, so that each symbol node in the TDG has a format of \texttt{\$name\$order(\$lineno)} to uniquely indicate a variable occurrence. For example, in Fig.~\ref{fig:mot_typegraph}, we create three symbol nodes (\texttt{pt0(9)}, \texttt{pt1(10), \texttt{pt2(12)}}) for variable \texttt{pt} as it appears three times in Listing~\ref{lst:mot_ex} (Line 9, 10, and 12).

    \begin{algorithm}[t]
    \caption{Type Dependency Graph Generation}
    \label{alg:typegraphgen}
    \begin{algorithmic}[1]
    \REQUIRE
    The AST of given function, $\mathit{func\_ast}$; \\
    \ENSURE
    Type dependency graph of the given function, $tg$\; \\
    
    Initialize an expression stack $\mathit{ex\_stack}$\; \\
    Initialize a variable stack $\mathit{var\_stack}$\; \\
    % $\mathit{ex\_stack}  \leftarrow$ new stack;   // expression stack
    % \STATE $var\_stack \leftarrow$  new stack  // variable stack
    \FORALL{$node \in$ func\_ast \&\& $\mathit{node}$ is not visited}
    \STATE // handle expression nodes
    \IF{$\mathit{node}$.type $\in$ Expressions} 
    \STATE $\mathit{ex\_stack}$.push($\mathit{node}$); $\mathit{ex\_node} \leftarrow$ new ex($\mathit{node}$)
    \STATE visit($\mathit{node}$.operands); $\mathit{ex\_stack}$.pop($\mathit{node}$)
    \IF{not $\mathit{ex\_stack}$.empty()}
    \STATE $tg$.addEdge($\mathit{ex\_node} \rightarrow \mathit{ex\_stack}$.top())
    \ENDIF
    \STATE $tg$.addNode($\mathit{ex\_node}$)
    \ENDIF
    \STATE // handle symbol nodes
    \IF{$\mathit{node}$.type == ast.Name}   
    \STATE $\mathit{sym\_node} \leftarrow$  new symbol($\mathit{node}$)
    \IF{$\mathit{node}$.ctx == write}
    \STATE $tg$.addEdge($\mathit{ex\_stack}$.top() $\rightarrow \mathit{sym\_node}$)
    \ELSE
    \STATE $tg$.addEdge($var\_stack$.top() $\rightarrow \mathit{sym\_node}$)
    \STATE $tg$.addEdge($\mathit{sym\_node} \rightarrow \mathit{ex\_stack}$.top())
    \ENDIF
    \STATE $var\_stack$.push($\mathit{sym\_node}$); $tg$.addNode($\mathit{sym\_node}$)
    \ENDIF
    \STATE // handle branch and merge nodes
    \IF{checkTypeBranch($\mathit{node}$)}   
    \STATE $\mathit{branch\_node} \leftarrow$ new branch($\mathit{node}$)
    \STATE $tg$.addNode($\mathit{branch\_node}$)
    \STATE $ctx1, ctx2 \leftarrow$ Branch($ctx$)
    \STATE visit($\mathit{node}$.left, $ctx1$); visit($\mathit{node}$.right, $ctx2$)
    \ENDIF
    \IF{checkTypeMerge($\mathit{node}$)} 
    \STATE $\mathit{merge\_node} \leftarrow$ new merge($\mathit{node}$)
    \STATE $tg$.addNode($\mathit{merge\_node}$)
    \STATE $ctx \leftarrow $ Merge($ctx1, ctx2$)
    \ENDIF
    \ENDFOR
    \end{algorithmic}
    \end{algorithm}

\textbf{Import Analysis.} Before establishing TDG for every input function, \tool first conducts import analysis to extract all user-defined types so that it can distinguish the initialization of user-defined types from regular function calls. \tool first collects all classes in source files, which constitute the initial set of user-defined types. Then it analyzes all local import statements such as ``\textit{from package import class}", and adds the imported classes into the user-defined type set. For all global import statements such as ``\textit{import package}", \tool locates the source of this package and adds all the classes and named tuples in the source into the user-defined type set. For each imported class, \tool solves the location of external source files and checks whether operator overloading methods exist in this class.

\textbf{Type Dependency Graph Generation.} 
Given the AST of input code and all the user-defined types extracted by import analysis, \tool creates TDG for each function based on the main logic shown in Alg.~\ref{alg:typegraphgen}. \tool first locates all the variables and expressions in the code by traversing the whole AST. Specifically, to visit each AST node, \tool employs the ASTVisitor provided by Python's module \textit{ast}~\cite{pyast}. \tool identifies expressions according to the definitions of expression nodes in Python (as depicted in Fig.~\ref{fig:expr}) and records every visited expression node using an expression stack. Whenever \tool identifies an expression node (Line 3), it builds a same node in the current TDG and pushes it into the expression stack. \tool will then recursively visits the expression's operands to capture new expression nodes until it encounters a variable node (Line 12), which is the leaf node of the AST.

\tool builds a symbol node in TDG for each visited identifier node of AST, and maintains a variable map to record all the occurrences of each variable. The AST already indicates the context of each variable occurrence, i.e., whether \textit{read} or \textit{write}.
\begin{enumerate}[leftmargin=*,wide, label=(\roman*)]
    \item If the variable context is \textit{read}, \tool will obtain the last occurrence of the variable according to the maintained variable map under the current context. It then creates an edge from the symbol node of the last occurrence to the symbol node of the current variable (Line 16 - 18).
    \item If the variable context is \textit{write}, \tool will fetch the value from the last expression in the expression stack and build an edge connecting from the expression node to the symbol node of the current variable (Line 14 - 15).
\end{enumerate}

Analogous to regular data flow analysis, \tool also checks whether the data flow branches (Line 23 - 27) or merges at certain locations (Line 29 - 32).

In TDG, each symbol node keeps a list of candidate types while each expression node includes type inference rules and type rejection rules. When \tool walks through TDG, the rules will be activated to produce new types. Thus, types can \textit{flow} from arguments to return values. By traversal, \tool obtains the types of each symbol node and outputs the type assignment. The leveraged type inference rules and type rejection rules are detailed in the next subsections.

\subsection{Static Type Inference}\label{sec:typeinfer}

\begin{figure*}[t]
\setlength{\belowcaptionskip}{-0.3cm}
\setlength{\abovecaptionskip}{0pt}
    \centering
    \begin{minipage}[t]{0.39\textwidth}
    \begin{align*}
    & \frac{v \in \textbf{Dom}(\pi)}{ \pi \vdash  v: \theta} \tag{Variable} \\
    & \frac{}{ \pi \vdash  c:\theta} \tag{Constant} \\
    & \frac{\pi \vdash e:\theta}{ \pi \vdash e.v:\theta.v} \tag{Attribute} \\
    & \frac{\pi \vdash e_1: \theta_1 \quad  \pi \vdash e_2: \theta_2 \quad \widetilde{\theta} = \{\textbf{bool}, \ \textbf{int}, \ O\}}{ \pi \vdash e_1 \ \textbf{bitop} \ e_2: \theta \wedge \widetilde{\theta} \quad \pi \vdash e_1: \theta_1 \wedge \widetilde{\theta} \quad \pi \vdash e_2: \theta_2 \wedge \widetilde{\theta}} \tag{LShift, RShift} \\
    & \pi \vdash e_1: \theta_1 \quad \widetilde{\theta} = \{\textbf{bool}, \ \textbf{int}, \ \textbf{float}, \ O\} \\
    & \frac{\pi \vdash e_2: \theta_2 \quad \theta' = getMorePreciseType(\theta_1 \wedge \widetilde{\theta}, \theta_2 \wedge \widetilde{\theta})}{ \pi \vdash e_1 \ \textbf{numop} \ e_2: \theta' \quad \pi \vdash \theta_1 \wedge \widetilde{\theta} \quad \pi \vdash \theta_2 \wedge \widetilde{\theta}} \tag{Numeric Operations} \\
    & \qquad \qquad \quad \pi \vdash e_1: \theta_1 \quad \widetilde{\theta_1} = \{\textbf{int}, \textbf{bool}\} \\
    & \frac{\pi \vdash e_2:\theta_2 \quad \widetilde{\theta_2} = \{\Gamma, \ \textbf{List}, \ \textbf{Tuple}, \ O\}}{ \pi \vdash e_1 \ \textbf{numop} \ e_2: \theta_2 \wedge \widetilde{\theta_2} \quad \pi \vdash e_1: \theta_1 \wedge \widetilde{\theta_1} \quad \pi \vdash e_2: \theta_2 \wedge \widetilde{\theta_2}} \tag{Mult} \\
    & \frac{\pi \vdash e_1:\theta_1 \quad  \pi \vdash e_2:\theta_2 \quad \widetilde{\theta} = \{\Gamma, \ \textbf{List}, \ \textbf{Tuple}, \ O\}}{ \pi \vdash e_1 \ \textbf{cmpop} \ e_2: \textbf{bool} \quad \pi \vdash e_1:\theta_1 \wedge \theta_2 \wedge \widetilde{\theta} \quad \pi \vdash e_2:\theta_1 \wedge \theta_2 \wedge \widetilde{\theta}} \tag{Lt,LtE,Gt,GtE} \\
    & \frac{}{ \pi \vdash u(e_1,...,e_n): u} \tag{Class Instantiation} \\
    & \frac{\pi \vdash e_1:\theta_1 \quad ... \quad \pi \vdash e_n:\theta_n}{\pi \vdash (e_1,...,e_n): \textbf{Tuple}[\theta _1,...,\theta _n] \quad \pi \vdash [e_1,...,e_n]: \textbf{List}[\theta _1,...,\theta _n]} \\ & \pi \vdash \{e_1,...,e_n\}: \textbf{Set}[\theta _1,...,\theta _n]  \tag{Tuple, List, Set} \\
    & \frac{\pi \vdash e:\theta \quad \widetilde{\theta} = \{A, \textbf{str}, \textbf{bytes}\}
     \quad \theta' = getElementType(\theta \wedge \widetilde{\theta})}{\pi \vdash \textbf{for } \ v \ \textbf{in} \ e:\theta' \quad \pi \vdash e: \theta \wedge \widetilde{\theta}} \tag{Comprehension} \\
    & \frac{\pi \vdash \textbf{for} \ v \ \textbf{in} \ e_1:\theta_1 \quad \pi \vdash e_2[v]: \theta_2}{\pi \vdash (e_2[v] \ \textbf{for} \ v \ \textbf{in} \ e_1): \textbf{Generator}[\theta_2]} \tag{Generator} \\
    & \frac{\pi \vdash \textbf{for} \ v \ \textbf{in} \ e_1:\theta_1 \quad \pi \vdash e_2[v]: \theta_2}{\pi \vdash [e_2[v] \ \textbf{for} \ v \ \textbf{in} \ e_1]: \textbf{List}[\theta_2]} \qquad \\ & \pi \vdash \{e_2[v] \ \textbf{for} \ v \ \textbf{in} \ e_1\}: \textbf{Set}[\theta_2] \tag{List, Set Comprehension}
    \end{align*}
    \end{minipage}
    \begin{minipage}[t]{0.39\textwidth}
    \begin{align*}
    & \frac{\pi \vdash e_1:\theta_1 \quad  \pi \vdash e_2:\theta_2}{ \pi \vdash e_1 \ \textbf{blop} \ e_2: \textbf{Union}[\theta_1, \theta_2]} \tag{Boolean Operations} \\
    & \frac{\pi \vdash e_1: \theta_1 \quad  \pi \vdash e_2: \theta_2 \quad \widetilde{\theta} = \{\textbf{bool}, \ \textbf{int}, \ \textbf{Set}, \ O\}}{ \pi \vdash e_1 \ \textbf{bitop} \ e_2: \theta \wedge \widetilde{\theta} \quad \pi \vdash e_1: \theta_1 \wedge \widetilde{\theta} \quad \pi \vdash e_2: \theta_2 \wedge \widetilde{\theta}} \tag{BitOr, BitAnd, BitXor} \\
    & \frac{\pi \vdash e_1: \theta_1 \quad  \pi \vdash e_2: \theta_2 \quad \widetilde{\theta} = \{\Gamma, \ \textbf{List}, \ \textbf{Tuple}, \ O\}}{ \pi \vdash e_1 \ \textbf{numop} \ e_2: \theta \wedge \widetilde{\theta} \quad \pi \vdash e_1:\theta_1 \wedge \theta_2 \wedge \widetilde{\theta} \quad \pi \vdash e_2:\theta_1 \wedge \theta_2 \wedge \widetilde{\theta} } \tag{Add} \\
    & \frac{\pi \vdash e_1: \theta_1 \quad  \pi \vdash e_2: \theta_2 \quad \widetilde{\theta} = \{\Gamma, \ \textbf{Set}, \ O\}}{ \pi \vdash e_1 \ \textbf{numop} \ e_2: \theta \wedge \widetilde{\theta} \quad \pi \vdash e_1:\theta_1 \wedge \theta_2 \wedge \widetilde{\theta} \quad \pi \vdash e_2:\theta_1 \wedge \theta_2 \wedge \widetilde{\theta}} \tag{Sub} \\
    & \frac{\pi \vdash e_1:\theta_1 \quad  \pi \vdash e_2:\theta_2}{ \pi \vdash e_1 \ \textbf{cmpop} \ e_2: \textbf{bool}} \tag{Eq,NotEq,Is,IsNot} \\
    &\pi \vdash e_1:\theta_1 \quad  \pi \vdash e_2:\theta_2 \\
    & \frac{\widetilde{\theta} = \{\textbf{str}, \textbf{bytes}, \textbf{List}, \ \textbf{Tuple}, \ \textbf{Set}, \ \textbf{Dict}, \ \textbf{Generator}\}}{ \pi \vdash e_1 \ \textbf{cmpop} \ e_2: \textbf{bool} \quad \pi \vdash e_2: \theta_2 \wedge \widetilde{\theta}} \tag{In,NotIn} \\
     &\pi \vdash e: \theta \quad \pi \vdash e_1:\theta_1 \quad ... \quad \pi \vdash e_n:\theta_n \\
    & \frac{\widetilde{\theta} =  \{\textbf{Callable}[[\theta_1,...,\theta_n],\theta]\} \quad \theta' = getReturnType(\theta \wedge \widetilde{\theta})}{ \pi \vdash e(e_1,...,e_n):\theta} \tag{Call} \\
    & \frac{\pi \vdash e_1:\theta_1 \quad ... \quad \pi \vdash e_n:\theta_n}{ \pi \vdash \{e_1:e_2,...,e_{n-1}:e_n\}: \textbf{Dict}[\theta_1:\theta_2,..., \theta _{n-1}:\theta_n]} \tag{Dict} \\ 
    & \frac{\pi \vdash e_1: \theta_1 \quad \pi \vdash e_2:\theta_2 \quad \widetilde{\theta_1} = \{\textbf{Dict}\} \quad \theta' = getValueType(\theta_1 \wedge \widetilde{\theta_1})}{ \pi \vdash e_1[e_2]:\theta' \quad \pi \vdash e_1:\theta_1 \wedge \widetilde{\theta_1}} \tag{SubScript} \\ 
    %& \frac{\Theta' = \Theta \wedge \{\textbf{Dict}\}}{\pi[e \mapsto \Theta] \rightarrow \pi [e \mapsto \Theta']} \tag{Bwd: SubScript} \\ 
    & \frac{\pi \vdash \textbf{for} \ v \ \textbf{in} \ e_1:\theta_1 \quad \pi \vdash e_2[v]: \theta_2 \quad \pi \vdash e_3[v]: \theta_3}{\pi \vdash \{e_2[v]:e_3[v] \ \textbf{for} \ v \ \textbf{in} \ e_1\}: \textbf{Dict}[\theta_2:\theta_3]} \tag{Dict Comprehension} \\
    & \pi \vdash e_1:\theta_1 \quad \pi \vdash e_2:\theta_2 \quad \widetilde{\theta_1} = \{A, \textbf{str}, \textbf{bytes}\} \\
    & \frac{\widetilde{\theta_2} = \{\textbf{int}, \textbf{bool}\} \quad \theta' = getElementType(\theta_1 \wedge \theta_2)}{ \pi \vdash e_1[e_2]:\theta' \quad \pi \vdash e_1: \theta_1 \wedge \widetilde{\theta_1} \quad \pi \vdash e_2: \theta_2 \wedge \widetilde{\theta_2}} \tag{Slice} \\
    %& \frac{\Theta' = \Theta \wedge \{\textbf{Tuple}, \ \textbf{List}, \ \textbf{Set}\ \textbf{str}, \ \textbf{bytes}\}}{\pi[e \mapsto \Theta] \rightarrow \pi [e \mapsto \Theta']} \tag{Bwd: Slice} \\
    \end{align*}
    \end{minipage}
\caption{Type inference and rejection rules of expressions in Python}
\label{fig:typingrule}
\end{figure*}

This section describes the type inference and rejection rules integrated in expression nodes, which are the key component of our static type inference. Fig.~\ref{fig:typingrule} denotes all the type inference and rejection rules used in static type inference. Each rule consists of some premises (contents above the line) and conclusions (contents below the line). They obey the following form: \\
\centerline{$\pi \vdash e:\theta\nonumber$.} \\
In this form, $\pi$ is called the context, which includes
lists that assign types to expression patterns. $e$ is the expression showed in Fig.~\ref{fig:expr}, and we use $e_1,...,e_n$ to represent different expressions. $\theta$ is the type showed in Fig.~\ref{fig:type}. We use  $\theta_1, ..., \theta_n$ to represent different types. A rule under this form is called a \textit{type judgement} or \textit{type assignment}. Our goal is to get the context $\pi$ that assigns types to all the variables in code. 

%Note that we do not consider statement types in the type rules, since \tool already involves the statement types into edges that connect different nodes in the TDG.

The premises of each rule in Fig.~\ref{fig:typingrule} are the types of input nodes $\theta_1, \theta_2, ...$ that constructs an expression, and the valid type set $\widetilde{\theta}$ for the current operation. 
Usually type inference rules only have one conclusion, which is the result type of current expression. However, as we also involve neural predictions in TDG and use type rejection rules to validate them, the conclusions of each rule in Fig.~\ref{fig:typingrule} have two parts: 1) the result type of current expression node and 2) the validated types of input nodes. (Some rules may not have the second part because they accept any input types.)

The result type of the current expression node is what traditional static type inference techniques usually infer. We denote it as \textit{forward type inference}. However, there exist types that are not allowed to conduct certain operations, which are guided by \textit{type constraints}. When a type constraint is violated, e.g., adding an integer to a string, traditional static inference techniques~\cite{mypy, pytype, pyre} throw type errors. For the wrongly predicted cases, \tool does not directly throw a type error since it accepts recommendations from DL models. To ``sanitize'' the recommendations from DL models, we create type rejection rules to validate and remove the wrong predictions in input nodes. We call this as \textit{backward type rejection}.

\textbf{Forward Type Inference.} 
\tool starts forward type inference with the nodes that do not have input nodes in TDG. It gradually visits all nodes in the graph and activates corresponding type inference rules if their premises are satisfied, i.e., all input nodes are fully inferred. This is the forward traversal of TDGs. As forward type inference in \tool is similar to traditional static type inference techniques, we only discuss the [Call] rule for which \tool has a special strategy. The premise of the [Call] rule requires the type of callees, which is beyond the scope of current functions. This premise is one major barrier for most static inference techniques to fully infer a program due to a large number of external APIs in Python programs~\cite{hu21static,pycg}. \tool only focuses on inferring the types of functions with explicit implementation in the current source code, in which the TDGs of the functions are connected.  \tool does not infer external calls for two reasons: 1) DL models perform well on predicting the types of commonly-used APIs that frequently occur in the training set; 2) Python maintains a \textit{typeshed}~\cite{typeshed} project to collect the type annotations of frequently-used modules, so \tool can directly access the types.

\textbf{Backward Type Rejection.} 
An input type in an expression must fulfill two constraints before it can conduct the expression: 1) it must be the valid type to conduct a certain expression, 2) it must have a valid relationship with other input types. \tool rejects the input types that violate these two constraints. It first checks whether the type is valid for an expression. We indicates valid types for each expression as $\widetilde{\theta}$ in Fig.~\ref{fig:typingrule}. For example, in [In, NotIn] rule, the types of $e_2$ must be iterable so \texttt{int} is not allowed and should not be in the valid type set $\widetilde{\theta}$. Then \tool checks whether the relationships between all inputs are valid. Apart from valid types for a certain operation, some operations also require the inputs to satisfy a certain relationship. For example, in [Add] rule, the two operands must have the same type. For types of two inputs \texttt{int} and \texttt{str}, even though they are in the valid type set of this operation, they are still rejected because they are not the same type. Therefore, in the [Add] rule, the final valid input types are the intersection of all input type sets $\theta_1$, $\theta_2$ and valid type set $\widetilde{\theta}$. 

Type Rejection rules can validate and reject the input types of an operation. However, the input types are the results of previous operations, so the type rejection process will also affect the input types of previous operations. To thoroughly remove the influence of wrong types, \tool also rejects the input types that result in the rejected types according to forward type inference rules. \tool gradually validates all type slots by starting from the type slots without output edges and producing the rejected input types. Then it traverses other slots until the whole TDG is visited. This is the backward traversal of TDGs.

    \begin{algorithm}[t]
    \caption{Type correction of user-defined types}
    \label{alg:typematch}
    \begin{algorithmic}[1]
    \REQUIRE
    Variable name, $name$; \\
    Valid user defined type set, $S$; \\
    Type String recommended by deep neural networks, $t$; \\
    Penalty added for name-type similarity to align with type-type similarity, $penalty$;
    \ENSURE
    Corrected type of current variable, $ct$;
    \IF {$t \in S$ or isBuiltin$(t)$}
    \STATE $ct \leftarrow t$;
    \ELSE
    \STATE $\text{largest\_sim} \leftarrow 0$; $\text{largest\_type} \leftarrow None$;
    \STATE $tw \leftarrow$ BPE$(t)$; $namew \leftarrow$ BPE$(name)$;
    \FOR{each $pt \in S$}
    \STATE $ptw \leftarrow$ BPE$(pt)$;
    \IF{$sim(ptw, tw) > \text{largest\_sim}$}
    \STATE $\text{largest\_sim} \leftarrow$ sim$(ptw, tw)$; $\text{largest\_type} \leftarrow pt$;
    \ENDIF
    \IF{$sim(ptw, namew) + penalty > \text{largest\_sim}$}
    \STATE $\text{largest\_sim} \leftarrow$ sim$(ptw, namew)$; $\text{largest\_type} \leftarrow pt$;
    \ENDIF
    \ENDFOR
    \STATE $ct \leftarrow \text{largest\_type}$;
    \ENDIF
    \end{algorithmic}
    \end{algorithm}

\textbf{Correctness.} Different from the DL-based approaches~\cite{deeptyper,mir2021type4py}, \tool can always guarantee the correctness  of its type assignments based on static inference. According to the architecture of \tool in Fig.~\ref{fig:arch}, the type assignments generated by \tool have two cases: 1) If the static inference can successfully handle a program, \tool does not need to invoke DL models to give type recommendations. Consequently, the type assignments fully based on the inference rules (Fig.~\ref{fig:typingrule}) are sound because they are collected from the Python official implementation CPython~\cite{cpython}; and 2) If the static inference cannot fully infer a program and the DL models are invoked to provide type recommendations (Sec. ~\ref{sec:typereco}), \tool utilizes type rejection rules to validate the recommendations and then calls the type inference rules again to infer the remaining types. In this case, our rejection rules thoroughly eliminate the influence of wrong recommendations, and the final results are also produced by static inference. Therefore, \tool always maintains the type correctness.

\eat{
[Variable] and [Constant] rules are two basic rules in the type system. \tool integrates all the types of literals beforehand so the context $\pi$ contains the type for each constant in the code. If a variable occurrence is previously typed, it is also included into $\pi$.

[Boolean Operations] can accept any types of the operands and the output type depends on the value of the input types. [Bitwise operations] are usually used on integers and the results are also integers. [Numeric operations] are usually used on numbers, including booleans, integers and floats, and the results are chosen from input types.
We use a predicate \textit{getMorePreciseType} to select the most precise type from all the input types of numbers. Generally, floating numbers are more precise than integers, and boolean can be used as an integer in Python. Besides, most built-in generic types such as \textit{list} overload the operators \textit{Mult}, \textit{Add} and \textit{Sub}. For the cases, the results will be the input generic types instead of numbers. Thus, we complement the cases with three separate rules ([Mult], [Add], [Sub]). [Compare operations] support most types and always return a boolean value.

[Tuple, List, Set, Dict] and [Tuple, List, Set, Dict Comprehensions] are two regular rules for constructing generic types. [Tuple, List, Set, Dict] initializes generic types and also their element types. Note that Python supports heterogeneous data structures, so a generic type can involve elements with different types, e.g., \texttt{List[int}, \texttt{str]}. This characteristic differs from most statically typed programming languages. Comprehension is an efficient way for users to create a generic type. To infer the type of elements generated from a comprehension, \tool uses the single [Comprehension] rule for predicting the parametric type $v$ of generic type $e$ within a $for$ $v$ $in$ $e$ statement.

[Call] handles all the internal function calls in code except for a special case, which is the $\_\_init\_\_()$ method of a class. It returns an instance of the class, and we use [Class Instantiation] rule to infer the type. The premise of the [Call] rule requires the type of callee, which is beyond the scope of current functions. This premise is one major barrier for most static inference techniques to fully infer a program, due to a large number of external APIs in Python programs. \tool only focuses on inferring the types of functions with explicit implementation in the current source code, in which the TDGs of the functions are connected.  \tool does not infer external calls for two reasons: 1) DL models perform well on predicting the types of commonly-used APIs as the APIs frequently occur in the training set; 2) Python maintains a \textit{typeshed}\footnote{\url{https://github.com/python/typeshed}} project to collect the type annotations of frequently-used modules, so \tool can easily access the types.

[Subscript] and [Slice] rules infer the types of elements from generic types, while [Attribute] rule infers the types of the attributes in classes.}

\subsection{Neural Type Recommendation}\label{sec:typereco}

\tool conducts static type inference based on type inference rules When static type inference can fully infer all the variables in TDG. However, some variables are hard to be statically typed so that \tool only gets a partially-inferred TDG. In this case, \tool asks DL models for recommendations. The neural type recommendation part of \tool includes two procedures: hot type slot identification and similarity-based type correction.

\begin{table*}[htb]
    \centering
    \caption{Comparison with the baseline approaches. Top-1,3,5 of \tool means it accepts 1,3,5 candidates from deep neural networks in type recommendation phase. The neural network in \tool is the corresponding comparison DL model.}
    \scalebox{0.85}{\begin{tabular}{clccc|cc|cc}
    \toprule
        \multirow{3}*{\textbf{Dataset}} &
         \multirow{3}*{\textbf{Type Category}}& \multirow{3}*{\textbf{Approach}}  & \multicolumn{2}{c}{\textbf{Top-1}} & \multicolumn{2}{c}{\textbf{Top-3}} & \multicolumn{2}{c}{\textbf{Top-5}}\\ 
         \cline{4-9} 
          &  & & \tabincell{c}{Exact \\ Match} & \tabincell{c}{Match to \\ Parametric} & \tabincell{c}{Exact \\ Match} & \tabincell{c}{Match to \\ Parametric} & \tabincell{c}{Exact \\ Match} & \tabincell{c}{Match to \\ Parametric} \\
         \hline 
         \hline
         \multirow{12}*{\tabincell{c}{\textbf{ManyTypes4Py}}} &
         \multirow{3}*{\textbf{Argument}} & Naive Baseline & 0.14 & 0.16 & 0.33 & 0.38 & 0.43 & 0.51 \\
          & & \typepy & 0.61 & 0.62 & 0.64 & 0.66 & 0.65 & 0.68  \\
         %\cmidrule{2-11}
        & & \tool & \textbf{0.65} & \textbf{0.67} & \textbf{0.70} & \textbf{0.74} & \textbf{0.72} & \textbf{0.76} \\
         \cline{2-9}
         & \multirow{3}*{\textbf{Return Value}} & Naive Baseline & 0.07 & 0.10 & 0.19 & 0.28 & 0.28 & 0.42 \\
         & & \typepy & 0.49 & 0.52 & 0.53 & 0.59 & 0.54 & 0.63   \\
         %\cmidrule{2-11}
         & & \tool & \textbf{0.60} & \textbf{0.72} & \textbf{0.63} & \textbf{0.76} & \textbf{0.65} & \textbf{0.77}  \\
         \cline{2-9}
         & \multirow{3}*{\textbf{Local Variable}} & Naive Baseline & 0.13 & 0.17 & 0.33 & 0.45 & 0.47 & 0.65 \\
         &  & \typepy & 0.67 & 0.73 & 0.71 & 0.78 & 0.72 & 0.79   \\
         & & \tool & \textbf{0.73} & \textbf{0.85} & \textbf{0.74} & \textbf{0.86} & \textbf{0.75} & \textbf{0.86} \\
         \cline{2-9}
         & \multirow{3}*{\textbf{All}} & Naive Baseline & 0.13 & 0.16 & 0.31 & 0.40 & 0.43 & 0.57 \\
         & & \typepy & 0.62 & 0.66 & 0.66 & 0.72 & 0.67 & 0.73   \\
         & & \tool & \textbf{0.69} & \textbf{0.77} & \textbf{0.72} & \textbf{0.81} & \textbf{0.72} & \textbf{0.82} \\
         \hline
         \hline
         \multirow{9}*{\tabincell{c}{\textbf{Typilus's} \\ \textbf{Dataset}}} &
         \multirow{3}*{\textbf{Argument}} & Naive Baseline & 0.19 & 0.20 & 0.38 & 0.42 & 0.46 & 0.50 \\
          & & \typilus & 0.60 & 0.65 & 0.69 & 0.74 & 0.71 & 0.76  \\
        & & \tool & \textbf{0.63} & \textbf{0.68} & \textbf{0.72} & \textbf{0.76} & \textbf{0.76} & \textbf{0.79} \\
         \cline{2-9}
         & \multirow{3}*{\textbf{Return Value}} & Naive Baseline & 0.11 & 0.11 & 0.28 & 0.31 & 0.36 & 0.43 \\
         & & \typilus & 0.41 & 0.57 & 0.48 & 0.62 & 0.50 & 0.64   \\
         %\cmidrule{2-11}
         & & \tool & \textbf{0.57} & \textbf{0.70} & \textbf{0.63} & \textbf{0.75} & \textbf{0.64} & \textbf{0.77}  \\
         \cline{2-9}
         & \multirow{3}*{\textbf{All}} & Naive Baseline & 0.17 & 0.18 & 0.35 & 0.39 & 0.44 & 0.48 \\
         & & \typilus & 0.54 & 0.62 & 0.63 & 0.70 & 0.65 & 0.72   \\
         %\cmidrule{2-11}
         & & \tool & \textbf{0.61} & \textbf{0.69} & \textbf{0.69} & \textbf{0.76} & \textbf{0.72} & \textbf{0.78}  \\

    \bottomrule
    \end{tabular}
    }
    %\vspace{-0.4cm}
    \label{tab:mainres}
\end{table*}

\textbf{Hot Type Slot Identification.} Some variables can impact the types of many other variables because they locate at the beginning of the data flow or possess type dependencies with many variables. We call these variables as \textit{hot type slots}. Given the types of hot type slots, static type inference techniques can infer the remaining type slots. Therefore, to optimize the type correctness of \tool, DL models are only invoked on the hot type slots instead of all the blank type slots.

To identify the hot type slots, \tool first removes slots already filled by static type inference and obtains a sub-graph with all the blank type slots. Then \tool employs a commonly-used dominator identification algorithm semi-NCA~\cite{seminca} to capture all dominators in the sub-graph. A node $X$ dominiating another node $Y$ in a graph means that each entry node to $Y$ must pass $X$. Thus, if a type slot $X$ dominates another type slot $Y$, $Y$'s type can be inferred from $X$'s type. \tool gradually removes the type slots $Y$ from the sub-graph until no type slots can be removed. In the smallest sub-graph, each type slot is not dominated by other type slots, and all the slots are hot type slots. For these type slots, \tool accepts type recommendations from DL models.

% After analyzing the partially inferred TDG, we find that a few variables can hinder the type inference of a large amount of variables because they locate at the beginning of data flow or have type dependencies with many variables. We call these variables \textit{hot type slots}, which means that we can still use static type inference to fully infer all variables once we know the types of these variables. To maintain the type correctness of \tool, we only asks DL models for recommendations on hot type slots instead of all blank type slots.

% To identify hot type slots from all the blank type slots, \tool first removes nodes that are already successfully inferred by static type inference and gets a sub-graph which contains all blank type slots. Then \tool uses a commonly used dominator identification algorithm semi-NCA~\cite{seminca} to find all dominators in this sub-graph. In graph theory, we say a node A dominates another node B, if every path from the entry node to B must go through A. Therefore, if a type slot A is dominated by type slot B, then its type can be inferred from B once we get the type of B. \tool removes node A from the sub-graph and will use static type inference to infer its type in the future. By gradually removing nodes, \tool finally gets the smallest sub-graph in which each node is not dominated by other nodes. The nodes in this smallest sub-graph is hot type slots. For types of these variables, \tool accepts recommendations from DL models.

\textbf{Similarity-based Type Correction for User-defined Types.} DL models provide one or more type recommendations for each hot type slot, depending on the strategy (Top-1, -3, or -5) \tool uses. Some DL models~\cite{deeptyper, typewriter} treat user-defined types as OOV tokens and do not predict the types, while other models~\cite{typilus, pytype} directly copy user-defined types from the training set but fail to predict those never appearing in the training set. We propose to complement the recommendation of user-defined types using the similarity-based type correction algorithm shown in Alg.~\ref{alg:typematch}. Note that \tool only focuses on replacing the explicitly incorrect user-defined types, i.e., those never imported or defined in current source file, with the most similar user-defined types collected by import analysis.

% \textbf{Similarity based user-defined type correction.} After identifying all hot type slots, \tool accepts recommendations from the top one or more types in the prediction list of the DL model (\textcolor{brown}{yellow} nodes in Fig.\ref{fig:arch}), depending on the strategy (Top-1, -3, or -5) \tool uses. As some DL models~\cite{deeptyper, typewriter} treat user-defined types as OOV tokens and do not predict such types while some models~\cite{typilus, pytype} copies user-defined types from training set but still can not handle those never appearing in the training set, in this phase \tool complement the recommendation of user-defined types by conducting a similarity based type correction algorithm shown in Algorithm~\ref{alg:typematch}. Note that \tool only tries to replace those explicitly incorrect user-defined types, i.e., those never imported or defined in current source file, with most similar user-defined types at this phase and leaves the accurate validation to static rules. 

Specifically, if the recommended type does not belong to built-in types, \tool checks whether the type appears in the user-defined type set collected from import analysis (Line 1). If the check result is False, the type will be regarded as explicitly incorrect and should be corrected. For these incorrect user-defined types, \tool replaces them with the most similar candidate in the user-defined type set. \tool employs Word2Vec~\cite{word2vec} to embed two types and the variable name into word embeddings, and calculates the cosine distance as the similarity of the two types (Line 6-12). For the OOV tokens, \tool splits them into subtokens using the BPE algorithm~\cite{BPE16,BPE94} (Line 5). Finally, \tool chooses the type candidate with the largest similarity to fill the user-defined type (Line 15).

\eat{

\subsection{Backward Type Rejection}\label{sec:typerej}
Type inference rules in Fig.~\ref{fig:typingrule} indicate how to conduct type inference given valid input types. However, there exist types that are not allowed to conduct some certain operations, which are guided by \textit{type constraints}. When a type constraint is violated, e.g., adding an integer to a string, traditional static inference techniques~\cite{mypy, pytype, pyre} throw type errors. For the wrongly-predicted cases, \tool will not directly throw a type error since it accepts recommendations from DL models. To ``sanitize'' the recommendations from DL models, we create type rejection rules to validate and remove the wrong predictions.

% how we can generate the type of result given valid input types. However, some types are not allowed to conduct some certain operations, which is guided by \textit{type constraints}. Traditional static inference techniques throw type errors when a type constraint is violated, e.g., when we try to add a integer to a string. As \tool accepts recommendations from DL models, it can not directly throw a type error when it encounters a wrong predicted type, instead it tries to "sanitize" the predictions. To achieve this goal, we create a new series of rules, called type rejection rules, to validate and remove wrong predictions from DL models.

\textbf{Generation of Rejected Types.} Fig.~\ref{fig:typerejrule} illustrates the type rejection rules which follow the similar form as the forward type inference but have the following differences: 1) We use the capitals of symbols in Fig.~\ref{fig:type} to indicate a set of types. For example, $\Theta$ indicates a set of candidate types for a variable, while $\theta$ only indicates single one of them in the type inference rules; 2) We use $\pi[e \mapsto \Theta] \mapsto \pi[e \mapsto \Theta']$ to indicate that the type of the expression $e$ changes from $\Theta$ to $\Theta'$ under context $\pi$.

% Fig.~\ref{fig:typerejrule} denotes how \tool validates and rejects wrong predictions from DL models. The rules described in Fig.~\ref{fig:typerejrule} follows the similar form in forward type inference but has the following differences: 1) We use the capital of symbols in Fig.~\ref{fig:type} to indicate a set of types. For example, $\Theta$ indicates all candidate types for a variable here while $\theta$ only indicates single one of them in typing rules. 2) We use $\pi[e \mapsto \Theta] \mapsto \pi[e \mapsto \Theta']$ to indicate that the types of expression $e$ changes from $\Theta$ to $\Theta'$ under context $\pi$.  

Type rejection rules verify and reject the input types of an operation. For example, given an operation \texttt{a + b}, where \texttt{a} has candidate types \texttt{int} and \texttt{str}, and \texttt{b} has only one candidate type \texttt{int}, type constraints of the operation \texttt{+} require the two operands to have the same type. So the type rejection rule [Add] filters out the wrong types by calculating the intersection of the two candidate type sets (e.g., in this case \texttt{int}) and removing all the types not in the intersection (e.g., \texttt{a}'s candidate type \texttt{str}). After validating the input types, the type rejection rule
% typing rule 
[Add] outputs the type as \texttt{int}.
%  can generate result type \texttt{int}.

\textbf{Backward Transmission of Rejected Types.} Type Rejection rules can validate and reject the input types of an operation. However, the input types are likely to be the results of previous operations, so the type removal will also affect the input types of previous operations. To handle the problem, \tool will also reject the input types that result in the rejected types. To gradually validate all the types in TDG, \tool starts from the slot without output edges and produces the rejected input types, and then traverses the other slots until the whole TDG is visited.}

% removing some of these types will also affect the input types of previous operations. To handle this problem, given the rejected result type, \tool finds the corresponding typing rule that generates this type and then also rejects the input types involving in this typing rule. To gradually validate all types in TDG, \tool starts from nodes that do not have output edges and generates its rejected input types, then it traverses the TDG until the whole TDG is visited. 

%\input{sections/setup}

\section{Evaluation}\label{sec:eval}
In the section, we answer the following research questions: \\
\textbf{RQ1:} How effective is \tool compared to baseline approaches? \\
\textbf{RQ2:} Can \tool well predict the rare types?\\
%How well does \tool handle the rare type prediction problem of DL approaches?\\
\textbf{RQ3:} What is the performance of the static type inference component in \tool?\\
% How well does \tool statically infer types without neural predictions compared to static type inference tools?\\

\vspace{-2em}

\subsection{Experimental Setup}

\begin{table}[t]
    \centering
    \caption{Type distribution in the test set. ``Rare'' indicates rare types and ``User'' indicates user-defined types.}
    \scalebox{0.81}{\begin{tabular}{ccc|cc|ccc}
    \toprule
         &\textbf{Category} & Total & Rare & \tabincell{l}{User} & Arg & \tabincell{l}{Return} & Local \\
         \hline
         \multirow{2}*{\tabincell{c}{\textbf{Typilus}}}&\textbf{Count} & 15,772 & 7,103 & 5,572  & 11,261  & 4,511 & -  \\
         \cline{2-8}
         &\textbf{Prop.} & 100\% & 45\% & 35\% & 71\% & 29\% & - \\
         \hline
         \hline
         \multirow{2}*{\textbf{Type4Py}}&\textbf{Count} & 37,408 & 14,035 & 10,023  & 11,807  & 5,491 & 20,110  \\
         \cline{2-8}
         &\textbf{Prop.} & 100\% & 37\% & 27\% & 32\% & 15\% & 53\% \\
         \bottomrule
         
    \end{tabular}
    }
    \vspace{-0.5cm}
    \label{tab:testset}
\end{table}

\textbf{Dataset.}
We used the two Python datasets mentioned in Sec.~\ref{sec:motivation} for evaluation. One is the \textit{Typilus's Dataset} released by Allamanis \etal~\cite{typilus}; and the other one is \textit{ManyTypes4Py} released by Mir \etal~\cite{type4pydataset}, with the number of different types in the test set and more detailed statistics shown in Table~\ref{tab:testset} and Sec.~\ref{sec:motivation}, respectively.
% We have introduced the details of these two datasets in Sec.~\ref{sec:motivation}. Table~\ref{tab:testset} shows the number of different types in the test set we use. 

\textbf{Evaluation Metrics.}
Following the previous work~\cite{typilus,mir2021type4py}, we choose two metrics \textit{Exact Match} and \textit{Match to Parametric} for evaluation. The two metrics compute the ratio of results that: 1) \textit{Exact Match}: completely matches human annotations. 2) \textit{Match to Parametric}: satisfy exact match when ignoring all the type parameters. For example, \texttt{List[int]} and \texttt{List[str]} are considered as matched under this metric.
% measure the type inference results by computing the ratio of results that satisfy: 1) \textit{Exact Match}: completely matches human annotations. 2) \textit{Match to Parametric}: Exact match when ignoring the all type parameters. For example, \texttt{List[int]} and \texttt{List[str]} are considered as matched under this metric.

\begin{table*}[htb]
    \centering
    \caption{\centering Comparison with the baseline DL approaches. 
    % \protect\\ (Top-1,3,5 of \tool means it accepts 1,3,5 candidates from deep neural networks in type recommendation phase)
    }
    \scalebox{0.85}{\begin{tabular}{clccc|cc|cc}
    \toprule
        \multirow{3}*{\textbf{Dataset}} &
         \multirow{3}*{\textbf{Type Category}}& \multirow{3}*{\textbf{Approach}}  & \multicolumn{2}{c}{\textbf{Top-1}} & \multicolumn{2}{c}{\textbf{Top-3}} & \multicolumn{2}{c}{\textbf{Top-5}}\\ 
         \cline{4-9} 
          &  & & \tabincell{c}{Exact \\ Match} & \tabincell{c}{Match to \\  Parametric} & \tabincell{c}{Exact \\ Match} & \tabincell{c}{Match to\\ Parametric} & \tabincell{c}{Exact \\ Match} & \tabincell{c}{Match to\\ Parametric} \\
         \hline 
         \hline
         \multirow{6}*{\tabincell{c}{\textbf{ManyTypes4Py}}} &
         \multirow{3}*{\textbf{User-defined Types}} & Naive Baseline & 0.00 & 0.00 & 0.00 & 0.00 & 0.00 & 0.00 \\
          & & \typepy & 0.29 & 0.29 & 0.34 & 0.34 & 0.36 & 0.36  \\
         %\cmidrule{2-11}
        & & \tool & \textbf{0.49} & \textbf{0.49} & \textbf{0.56} & \textbf{0.56} & \textbf{0.58} & \textbf{0.58} \\
         \cmidrule{2-9}
         & \multirow{3}*{\textbf{Rare Types}} & Naive Baseline & 0.03 & 0.07 & 0.08 & 0.21 & 0.13 & 0.35 \\
         & & \typepy & 0.39 & 0.46 & 0.45 & 0.54 & 0.47 & 0.57   \\
         %\cmidrule{2-11}
         & & \tool & \textbf{0.51} & \textbf{0.66} & \textbf{0.56} & \textbf{0.72} & \textbf{0.58} & \textbf{0.73}  \\
         \hline
         \hline
         \multirow{6}*{\tabincell{c}{\textbf{Typilus's} \\ \textbf{Dataset}}} &
         \multirow{3}*{\textbf{User-defined Types}} & Naive Baseline & 0.00 & 0.00 & 0.00 & 0.00 & 0.00 & 0.00 \\
          & & \typilus & 0.32 & 0.32 & 0.40 & 0.40 & 0.42 & 0.42  \\
        & & \tool & \textbf{0.47} & \textbf{0.47} & \textbf{0.56} & \textbf{0.56} & \textbf{0.60} & \textbf{0.60} \\
         \cline{2-9}
         & \multirow{3}*{\textbf{Rare Types}} & Naive Baseline & 0.00 & 0.01 & 0.01 & 0.03 & 0.03 & 0.09 \\
         & & \typilus & 0.32 & 0.43 & 0.41 & 0.53 & 0.43 & 0.55   \\
         %\cmidrule{2-11}
         & & \tool & \textbf{0.43} & \textbf{0.55} & \textbf{0.52} & \textbf{0.63} & \textbf{0.56} & \textbf{0.67}  \\

    \bottomrule
    \end{tabular}
    }
    %\vspace{-0.4cm}
    \label{tab:detailedres}
\end{table*}

\textbf{Baseline Approaches.}
To verify the effectiveness of the proposed \tool, we choose five baseline approaches for comparison:

1) A naive baseline. It represents a basic data-driven method. We build this baseline following the work~\cite{typewriter}, which makes predictions by sampling form the distribution of the most frequent ten types.

2) \pytype~\cite{pytype} and \pyre~\cite{pyre}. They are two popular Python static type inference tools from Google and Facebook, respectively. 

3) \typilus~\cite{typilus} and Type4Py~\cite{mir2021type4py}.  \typilus is a graph model that utilizes code structural information. Type4Py is a hierarchical neural network that uses type clusters to predict types.

\textbf{Implementation of \tool}
The entire framework of \tool is implemented using Python, which contains more than 9,000 lines of code. We obtain all typing rules and rejection rules from Python's official documentation~\cite{py3doc} and its implementation CPython\footnote{https://github.com/python/cpython}. We use Word2Vec model from the gensim library~\cite{gensim} as the embedding when calculating the similarity between two types. We train the Word2Vec model by utilizing all the class names and variable names in the training set of \typilus. The dimension of the word embeddings and size of the context window are set as 256 and 10, respectively. Due to the small training corpus for Word2Vec, we choose Skip-Gram algorithm for model training~\cite{DBLP:conf/nips/MikolovSCCD13}. We choose \typilus and Type4Py as the neural network model from which \tool accepts type recommendations. We choose the exact hyper-parameters for \typilus and Type4Py used in the original papers. We run all experiments on Ubuntu 18.04. The system has a Intel(R) Xeon(R) CPU (@2.4GHz) with 32GB RAM and 2 NVIDIA TiTAN V GPUs with 12GB RAM.

\subsection{RQ 1: Effectiveness of \tool}
%  compared to DL approaches
% We evaluate \tool and other baseline approaches considering different type categories, including arguments, local variables and return values. Table~\ref{tab:mainres} shows the evaluation results. 
We evaluate the effectiveness of \tool considering different type categories, including arguments, local variables, and return values. The results are depicted in Table~\ref{tab:mainres}.

\textbf{Overall performance.} The naive baseline achieves high scores regarding the top-5 exact match metric for different type categories, some of which are even close to the performance of DL models. Since the naive baseline only predicts types with high occurrence frequencies in the dataset, the results indicate the challenge of accurately predicting rare types. \typilus and Type4Py mitigate the challenge by using similarity learning and type clusters and achieve $\sim$0.6 regarding the top-1 exact match metric. \tool further improves the metric by 11\% and 15\% compared with \typilus and Type4Py, respectively. \tool also enhances the top-1 match to parametric metric by 17\% and 11\% compared with \typilus and Type4Py, respectively. The improvement indicates the effectiveness of \tool in accurate type prediction. Besides, \tool presents better performance than the respective DL models regarding the top-3,5 metrics, demonstrating that \tool infers new results based on the static type inference rules, instead of just filtering out or reordering the predictions of DL models.

\textbf{Type categories.} Both Type4Py and \typilus perform better on the argument category than the return value category, which may reflect the difficulty of predicting the types of return values. By building upon type inference rules and TDGs, \tool can handle the complicated type dependencies of return values and thereby improve Type4Py and \typilus by 22\% and 39\%, respectively, w.r.t. the Top-1 exact match metric. \tool also slightly meliorates the prediction of the argument category by 7\% and 5\% compared with Type4Py and \typilus, respectively. The improvement may be attributed to the type correction for user-defined types. Moreover, \tool outperforms Type4Py by 9\% for predicting local variables.

\begin{tcolorbox}[breakable,width=\linewidth,boxrule=0pt,top=1pt, bottom=1pt, left=1pt,right=1pt, colback=gray!20,colframe=gray!20]
\textbf{Answer to RQ1:} \tool shows great improvement (11\% $\sim$ 15\%) on overall type inference performance, and the most significant improvement is on return value inference (22\% $\sim$ 39\%).
\end{tcolorbox}

\vspace{-1em}

% For specific type categories, both Type4Py and \typilus achieve better performance on arguments than return values. This indicates the difficulty of predicting types of return values. With supports of typing rules and type dependency graphs, \tool can handles the complicated type dependencies of return values thus it improves Type4Py's performance by 22\% and Typilus's performance by 39\% on return values. It balances the performance of arguments and return values so that it does not suffer from the problem of predicting any specific type category. \tool also slightly improves the performance on arguments by 7\% compared with Type4Py and 5\% compared with \typilus. This contributes to the user-defined type correction algorithm and call analysis it uses. Type4Py achieves the best performance on local variables. This may because local variables contains some type dependencies that are not very complicated and can be well handled by DL models while arguments almost have few type dependencies and most return values have complicated type dependencies. \tool also improves Type4Py's performance by 9\% on local variables.

\subsection{RQ 2: Prediction of Rare Types}
% How well does \tool handle the rare type prediction problem of DL approaches?
Rare types are defined as the types with proportions less than 0.1\% among the annotations in the datasets, and we observe that 99.7\% and 79.0\% of rare types are user-defined types in ManyTypes4Py and Typilus's dataset, respectively. Table~\ref{tab:detailedres} illustrates the prediction results of rare types and user-defined types. We can observe that the naive baseline barely infers rare types and user-defined types. Besides, the performance of Type4Py and \typilus drops significantly for the two type categories, which indicates that type occurrence frequencies can impact the performance of DL models. \tool shows the best performance on predicting the two type categories. Specifically, for inferring the rare types, \tool outperforms Type4Py and Typilus by 31\% and 34\%, respectively, w.r.t. the top-1 exact match metric. Regarding the prediction of user-defined types, \tool increases the performance of Type4Py and \typilus by 69\% and 47\%, respectively.

\begin{tcolorbox}[breakable,width=\linewidth,boxrule=0pt,top=1pt, bottom=1pt, left=1pt,right=1pt, colback=gray!20,colframe=gray!20]
\textbf{Answer to RQ2:} \tool greatly alleviates the prediction issue of rare types faced by DL models by achieving a $>30$\% boost, taking the advantage of the static type inference component.
\end{tcolorbox}

\vspace{-1em}

\subsection{RQ 3: Performance of the Static Type Inference Component}
% How well does \tool statically infer types without neural predictions compared to static type inference tools?

\begin{table}[htb]
    \centering
    \setlength{\belowcaptionskip}{-0.4cm}
    \caption{\centering Comparison with static type inference tools.}
    \begin{threeparttable}
    \scalebox{0.83}{\begin{tabular}{clccr}
    \toprule
        \multirow{2}*{\textbf{Dataset}} &
         \multirow{2}*{\textbf{\tabincell{c}{Type \\ Category}}}& \multirow{2}*{\textbf{Approach}}  &
         \tabincell{c}{\textbf{Exact}} &
         \tabincell{c}{\textbf{\#Correct}} \\
         & & & \textbf{Match} & \textbf{Annotations} \\
         \hline 
         \hline
         \multirow{9}*{\tabincell{c}{\textbf{ManyTypes4Py} }} &
         \multirow{3}*{\textbf{Argument}} & \pytype & - & 0 \\
          & & \pyre & \textbf{0.96} & 613 \\
        & & \tool & 0.94 & \textbf{1060} \\
        \cline{2-5}
         & \multirow{3}*{\textbf{\tabincell{l}{Return \\ Value}}} & \pytype & 0.81 & 777 \\
         & & \pyre & 0.84 & 662  \\
         & & \tool & \textbf{0.86} & \textbf{2603}  \\
         \cline{2-5}
         & \multirow{3}*{\textbf{All}} & \pytype & 0.81 & 777  \\
         & & \pyre & \textbf{0.89} & 1275   \\
         & & \tool & 0.88 & \textbf{3663 (16918*)} \\
         \hline
         \hline
         \multirow{9}*{\tabincell{c}{\textbf{Typilus's} \\ \textbf{Dataset}}} &
         \multirow{3}*{\textbf{Argument}} & \pytype & - & 0 \\
          & & \pyre & \textbf{0.96} & 543  \\
        & & \tool & 0.88 & \textbf{983} \\
         \cline{2-5}
         & \multirow{3}*{\textbf{\tabincell{l}{Return \\ Value}}} & \pytype & 0.79 & 552 \\
         & & \pyre & 0.71 & 484   \\
         & & \tool & \textbf{0.91} & \textbf{2461}  \\
         \cline{2-5}
         & \multirow{3}*{\textbf{All}} & \pytype & 0.79 & 552 \\
         & & \pyre & 0.82 & 1027  \\
         & & \tool & \textbf{0.90} & \textbf{3444}  \\

    \bottomrule
    \end{tabular}
    }
    \begin{tablenotes}\footnotesize
    \item[*] The number of correct annotations when including local variables.
    \end{tablenotes}
    \end{threeparttable}
    \label{tab:staticres}
    %\vspace{-0.3cm}
\end{table}
% \footnotetext[1]{The number of correct annotations when including local variables.}
In this RQ, we evaluate the performance of the static type inference component in \tool compared with popular static type inference tools Pytype~\cite{pytype} and Pyre~\cite{pyre}. The results are shown in Table~\ref{tab:staticres}. We only consider the type categories of argument and return value for comparison since Pyre and Pytype do not infer types for local variables. We use the metric \textit{number of correct annotations} to replace the metric \textit{match to parametric} that is usually used to evaluate DL models, considering that the results of static inference are exact and not recommendations.

As shown in Table~\ref{tab:staticres}, the exact match scores of all the static tools are greatly high, and \tool achieves the best performance. The results indicate the effectiveness of the static type inference component in \tool. We also find that there remains $\sim$10\% of the results inconsistent with human annotations in the datasets. By using Python's official type checker \textit{mypy} to check these results, we observe that all the types annotated by \tool do not produce type errors, which reflects the correctness of the proposed \tool. After manual checking of these inconsistent types, we find this inconsistency is caused by subtypes, we further discuss them in Sec.~\ref{sec:discussion}. Besides, \textit{mypy}'s results indicate very few inconsistent cases are caused by incorrect human annotations. To test whether \tool can rectify the incorrect annotations, we replace the original annotations with the results inferred by \tool, and inspect whether the original type errors are fixed. We finally correct 7 annotations on 6 GitHub repositories, including memsource-wrap~\cite{memsource-wrap}, MatasanoCrypto~\cite{MatasanoCrypto}, metadata-check~\cite{metadata-check}, coach~\cite{coach}, cauldron~\cite{cauldron}, growser~\cite{growser}, and submit pull requests to these repository owners. The owners of MatasanoCrypto and cauldron have approved our corrections.

While Pytype and Pyre present high exact match scores, the numbers of variables they can accurately infer are small. Table~\ref{tab:staticres} shows that \tool generally outputs 2x argument types and 3x return value types compared with them in both datasets, which suggests \tool's stronger inference ability than Pyre and Pytype. Such improvements attribute to \tool's import analysis and [Class Instantiation] rule on supporting the inference of user-defined types, and inter-procedural analysis on supporting the inference of class attributes and functions.

\begin{tcolorbox}[breakable,width=\linewidth,boxrule=0pt,top=1pt, bottom=1pt, left=1pt,right=1pt, colback=gray!20,colframe=gray!20]
\textbf{Answer to RQ3:} Only considering the static inference part, \tool still outperforms current static type inference tools by inferring 2$\times$ $\sim$ 3$\times$ more variables with higher accuracy.
\end{tcolorbox}

\vspace{-1em}

\section{Discussion}\label{sec:discussion}
\textbf{Inference of subtypes.} Although \tool achieves promising results for type prediction and passes the check of \textit{mypy}, it is still unable to infer some variable types (around 10\%). The failure mainly occurs in the inference of subtypes.
\begin{lstlisting}[language = python,caption = An example HiTyper fails to infer., label = lst:failcase]
#File: miyakogi.wdom/wdom/node.py
#Human annotation: AbstractNode
#Typilus: ForeachClauseNode      HiTyper: Node
def _append_element(self, node: AbstractNode) -> AbstractNode:
    if node.parentNode:
        node.parentNode.removeChild(node)
    self.__children.append(node)
    node.__parent = self
    return node
def _append_child(self, node):
    if not isinstance(node, Node):
        raise TypeError
    ...
    return self._append_element(node)
\end{lstlisting}

Listing~\ref{lst:failcase} shows an example for which \tool's result is inconsistent with the original annotations but still passes the check of \textit{mypy}. The return statement at Line 9 indicates that the type of return value is the same as the type of argument \texttt{node}. \typilus predicts the type as \texttt{ForeachClauseNode}, which is invalid since it is not imported in the code and is from other projects in the training set. \tool infers the type as \texttt{xml.dom.Node}, because the function is called by another function named $\_append\_child$ in the same file and the caller transmits a variable with type \texttt{Node}. However, developers annotate the variable as \texttt{AbstractNode}, the parent type. Such behavior is common in practice and poses a challenge for accurate type prediction.

\section{Related Work}\label{sec:literature}

% Recently, type inference of dynamic programming languages becomes an increasingly hot topic. It is primarily used to enhance the type safety of programs and reduce type errors. Since nowadays the type information of variables can be beneficial for capturing the semantics of a program~\cite{Allamanis18graph}, detect bugs~\cite{Gao17type} and recommend APIs~\cite{he2021pyart}, the granularity of the type inference task evolve from function/API level to variable level.

\textbf{Static and dynamic type inference.} Existing static type inference techniques towards different programming languages, such as Java~\cite{javainfer05}, JavaScript~\cite{jsinfer09}, Ruby~\cite{rubyinfer09}, Python~\cite{maxsmtpython18} or using different static analysis techniques~\cite{emmi16abstracttypeinfer,Zvonimir21dataflowtypeinfer,sheng16principaltypeinfer}, and inference tools used in industry such as Pytype~\cite{pytype}, Pysonar2~\cite{pysonar} and Pyre~\cite{pyre} are correct by design with relatively high accuracy on some simple builtin types and generic types, but due to the dynamic feature~\cite{jsdynamic} of programming languages, they can hardly handle user-defined types and some complicated generic types. \tool extends the inference ability of static inference techniques by conducting import analysis and inter-procedural analysis to handle the user-defined types, class attributes and functions in code. Dynamic type inference techniques~\cite{rubydinfer11,rubychecker13, Yusuke19dynamictypeinfer} and type checkers such as Mypy~\cite{mypy}, Pytype~\cite{pytype}, Pyre Check~\cite{pyre}, Pyright~\cite{pyright} calculate the data flow between functions and infer types according to several input cases. They can more accurately predict types than static type inference techniques but have limited code coverage and large time consumption. Thus, they encounter difficulties when deployed on large scales of code.

\textbf{Machine learning in type inference.}
Traditional static and dynamic type inference techniques employ rule-based methods and give the exact predicted type for each type slot. Xu \etal~\cite{pbinfer16} introduce probabilistic type inference, which returns several candidate types for one variable. Hellendoorn \etal~\cite{deeptyper} regard types as word labels and build a sequence model DeepTyper to infer types. However, their model treats each variable occurrence as a new variable without strict constraints. Dash \etal~\cite{dash18concepttype} introduce conceptual types which divide a single type such as \texttt{str} to more detailed types such as \texttt{url},\texttt{phone}, etc. Pradel \etal~\cite{typewriter} design 4 separate sequence models to infer function types in Python. They also add a validation phase to filter out most wrong predictions using type checkers. Allamanis \etal~\cite{typilus} propose a graph model to represent code and use KNN to predict the types. The method enlarges type set but still fails when the predicted types are not occurring in the training set. Although DL models have shown great improvement in this task, it still faces the type correctness and rare type prediction problem, \tool addresses these two problems by integrating DL models into the framework of static inference since static inference is data-insensitive and implemented on type inference rules that are sound by design. Despite efforts on Python type inference, there are also a bunch of work on type inference of other dynamically typed programming languages. Wei \etal~\cite{wei2020lambdanet} propose a neural graph network named LambdaNet to conduct probabilistic type inference on JavaScript programs. Jesse \etal~\cite{jesse2021typebert} propose a BERT-style model named TypeBert that obtains better performance on type inference of JavaScript than most sophisticated models.

\section{Conclusion}\label{sec:con}
In the work, we propose \tool, a hybrid type inference framework which iteratively integrates DL models and static analysis for type inference. \tool creates TDG for each function and validates predictions from DL models based on typing rules and type rejection rules. Experiments demonstrate the effectiveness of \tool in type inference, enhancement for predicting rare types, and advantage of the static type inference component in \tool. %\tool is open-sourced at \url{https://github.com/JohnnyPeng18/HiTyper}.

% We present \tool, a static type inference framework with neural predictions to infer types in Python source code. \tool leverages type dependency graph, typing rules and type rejection rules to validate predictions from DL models and mitigate the rare type prediction problem of DL models. Experiments show that \tool can generally and significantly improve the performance of \typilus and Type4Py.

\section{Acknowledgments}
The authors would like to thank the efforts made by anonymous reviewers. The work described in this paper was supported by the Research Grants Council of the Hong Kong Special Administrative Region, China (No. CUHK 14210920 of the General Research Fund), National Natural Science Foundation of China under project No. 62002084, Stable support plan for colleges and universities in Shenzhen under project No. GXWD20201230155427003-20200730101839009, and supported, in part, by Amazon under an Amazon Research Award in automated reasoning; by the United
States \grantsponsor{GS100000001}{National Science Foundation}{https://www.nsf.gov/} (NSF) under grants No. ~\grantnum{GS100000001}{1917924} and No.~\grantnum{GS100000001}{2114627}; and by the  \grantsponsor{GS100000002}{Defense Advanced Research Projects Agency}{https://www.darpa.mil/}  (DARPA) under grant~\grantnum{GS100000002}{N66001-21-C-4024}. 
Any opinions, findings, and conclusions or recommendations expressed in this publication are those of the authors, and do not necessarily reflect the views of the above sponsoring entities.

% \section{Appendices}

\balance

\bibliographystyle{plain}
\bibliography{ref}

\begin{thebibliography}{10}

\bibitem{MatasanoCrypto}
aldur.
\newblock The return value at line 295., 2021.
\newblock
  https://github.com/aldur/MatasanoCrypto/blob/master/matasano/blocks.py.

\bibitem{typilus}
Miltiadis Allamanis, Earl~T. Barr, Soline Ducousso, and Zheng Gao.
\newblock Typilus: Neural type hints.
\newblock In {\em Proceedings of the 41st ACM SIGPLAN Conference on Programming
  Language Design and Implementation}, PLDI 2020, page 91–105, New York, NY,
  USA, 2020. Association for Computing Machinery.

\bibitem{rubydinfer11}
Jong-hoon~(David) An, Avik Chaudhuri, Jeffrey~S. Foster, and Michael Hicks.
\newblock Dynamic inference of static types for ruby.
\newblock {\em SIGPLAN Not.}, 46(1):459–472, January 2011.

\bibitem{javainfer05}
Christopher Anderson, Paola Giannini, and Sophia Drossopoulou.
\newblock Towards type inference for javascript.
\newblock In {\em Proceedings of the 19th European Conference on
  Object-Oriented Programming}, ECOOP'05, page 428–452, Berlin, Heidelberg,
  2005. Springer-Verlag.

\bibitem{sheng16principaltypeinfer}
Sheng Chen and Martin Erwig.
\newblock Principal type inference for gadts.
\newblock In Rastislav Bod{\'{\i}}k and Rupak Majumdar, editors, {\em
  Proceedings of the 43rd Annual {ACM} {SIGPLAN-SIGACT} Symposium on Principles
  of Programming Languages, {POPL} 2016, St. Petersburg, FL, USA, January 20 -
  22, 2016}, pages 416--428. {ACM}, 2016.

\bibitem{dash18concepttype}
Santanu~Kumar Dash, Miltiadis Allamanis, and Earl~T. Barr.
\newblock Refinym: Using names to refine types.
\newblock In {\em Proceedings of the 2018 26th ACM Joint Meeting on European
  Software Engineering Conference and Symposium on the Foundations of Software
  Engineering}, ESEC/FSE 2018, page 107–117, New York, NY, USA, 2018.
  Association for Computing Machinery.

\bibitem{emmi16abstracttypeinfer}
Michael Emmi and Constantin Enea.
\newblock Symbolic abstract data type inference.
\newblock In Rastislav Bod{\'{\i}}k and Rupak Majumdar, editors, {\em
  Proceedings of the 43rd Annual {ACM} {SIGPLAN-SIGACT} Symposium on Principles
  of Programming Languages, {POPL} 2016, St. Petersburg, FL, USA, January 20 -
  22, 2016}, pages 513--525. {ACM}, 2016.

\bibitem{mypydoc}
Python~Software Foundation.
\newblock Official documentation of {Mypy}, 2020.
\newblock https://mypy.readthedocs.io/en/stable/builtin\_types.html.

\bibitem{py3doc}
Python~Software Foundation.
\newblock Official documentation of {Python3}, 2020.
\newblock https://docs.python.org/3.

\bibitem{cpython}
Python~Software Foundation.
\newblock Cpython. python's official implementation, 2021.
\newblock https://github.com/python/cpython.

\bibitem{rubyinfer09}
Michael Furr, Jong-hoon~(David) An, Jeffrey~S. Foster, and Michael Hicks.
\newblock Static type inference for ruby.
\newblock In {\em Proceedings of the 2009 ACM Symposium on Applied Computing},
  SAC '09, page 1859–1866, New York, NY, USA, 2009. Association for Computing
  Machinery.

\bibitem{BPE94}
Philip Gage.
\newblock A new algorithm for data compression.
\newblock {\em C Users J.}, 12(2):23–38, February 1994.

\bibitem{memsource-wrap}
gengo.
\newblock The return value at line 853., 2021.
\newblock https://github.com/gengo/memsource-wrap/blob/master/memsource/api.py.

\bibitem{seminca}
Loukas Georgiadis, Robert Tarjan, and Renato Werneck.
\newblock Finding dominators in practice.
\newblock volume~10, pages 69--94, 01 2006.

\bibitem{octoverse}
Github.
\newblock The 2020 state of the octoverse, 2020.
\newblock https://octoverse.github.com/.

\bibitem{Hanenberg13study}
Stefan Hanenberg, Sebastian Kleinschmager, Romain Robbes, Éric Tanter, and
  Andreas Stefik.
\newblock An empirical study on the impact of static typing on software
  maintainability.
\newblock {\em Empirical Software Engineering}, 19, 10 2013.

\bibitem{maxsmtpython18}
Mostafa Hassan, Caterina Urban, Marco Eilers, and Peter M{\"u}ller.
\newblock Maxsmt-based type inference for python 3.
\newblock In Hana Chockler and Georg Weissenbacher, editors, {\em Computer
  Aided Verification}, pages 12--19, Cham, 2018. Springer International
  Publishing.

\bibitem{deeptyper}
Vincent~J. Hellendoorn, Christian Bird, Earl~T. Barr, and Miltiadis Allamanis.
\newblock Deep learning type inference.
\newblock In {\em Proceedings of the 2018 26th ACM Joint Meeting on European
  Software Engineering Conference and Symposium on the Foundations of Software
  Engineering}, ESEC/FSE 2018, page 152–162, New York, NY, USA, 2018.
  Association for Computing Machinery.

\bibitem{hu21static}
Mingzhe Hu, Yu~Zhang, Wenchao Huang, and Yan Xiong.
\newblock Static type inference for foreign functions of python.
\newblock In {\em 32nd International Symposium on Software Reliability
  Engineering}, October 2021.

\bibitem{coach}
IntelLabs.
\newblock The return value at line 95.
\newblock
  \url{https://github.com/IntelLabs/coach/blob/master/rl\_coach/memories/non\_episodic/experience\_replay.py},
  2021.

\bibitem{jsinfer09}
Simon~Holm Jensen, Anders M\o{}ller, and Peter Thiemann.
\newblock Type analysis for javascript.
\newblock In {\em Proceedings of the 16th International Symposium on Static
  Analysis}, SAS '09, page 238–255, Berlin, Heidelberg, 2009.
  Springer-Verlag.

\bibitem{jesse2021typebert}
Kevin Jesse, Premkumar~T. Devanbu, and Toufique Ahmed.
\newblock Learning type annotation: Is big data enough?
\newblock In {\em Proceedings of the 29th ACM Joint Meeting on European
  Software Engineering Conference and Symposium on the Foundations of Software
  Engineering}, ESEC/FSE 2021, page 1483–1486, New York, NY, USA, 2021.
  Association for Computing Machinery.

\bibitem{jetbrainssurvey}
Jetbrains.
\newblock Python developer survey conducted by jetbrains and python software
  foundation, 2020.
\newblock https://www.jetbrains.com/lp/python-developers-survey-2020/.

\bibitem{kang2020decoupling}
Bingyi Kang, Saining Xie, Marcus Rohrbach, Zhicheng Yan, Albert Gordo, Jiashi
  Feng, and Yannis Kalantidis.
\newblock Decoupling representation and classifier for long-tailed recognition,
  2020.

\bibitem{typesurvey}
C.~M. {Khaled Saifullah}, M.~{Asaduzzaman}, and C.~K. {Roy}.
\newblock Exploring type inference techniques of dynamically typed languages.
\newblock In {\em 2020 IEEE 27th International Conference on Software Analysis,
  Evolution and Reengineering (SANER)}, pages 70--80, 2020.

\bibitem{numba}
Siu~Kwan Lam, Antoine Pitrou, and Stanley Seibert.
\newblock {Numba}: A {LLVM}-based {Python} {JIT} compiler.
\newblock In {\em 2nd LLVM Workshop on the LLVM Compiler Infrastructure in
  HPC}.

\bibitem{Le20survey}
Triet H.~M. Le, Hao Chen, and Muhammad~Ali Babar.
\newblock Deep learning for source code modeling and generation: Models,
  applications, and challenges.
\newblock {\em ACM Comput. Surv.}, 53(3), June 2020.

\bibitem{pep589}
Jukka Lehtosalo.
\newblock {PEP} 589 -- {TypedDict}: Type hints for dictionaries with a fixed
  set of keys, March 2019.
\newblock https://www.python.org/dev/peps/pep-0589/.

\bibitem{pep544}
Ivan Levkivskyi, Jukka Lehtosalo, and Łukasz Langa.
\newblock {PEP} 544 -- protocols: Structural subtyping (static duck typing),
  March 2017.
\newblock https://www.python.org/dev/peps/pep-0544/.

\bibitem{liu2020deep}
Jialun Liu, Yifan Sun, Chuchu Han, Zhaopeng Dou, and Wenhui Li.
\newblock Deep representation learning on long-tailed data: A learnable
  embedding augmentation perspective, 2020.

\bibitem{NL2Type}
Rabee~Sohail Malik, Jibesh Patra, and Michael Pradel.
\newblock Nl2type: Inferring javascript function types from natural language
  information.
\newblock In {\em Proceedings of the 41st International Conference on Software
  Engineering}, ICSE '19, page 304–315. IEEE Press, 2019.

\bibitem{word2vec}
Tomas Mikolov, Ilya Sutskever, Kai Chen, Greg Corrado, and Jeffrey Dean.
\newblock Distributed representations of words and phrases and their
  compositionality.
\newblock NIPS'13, page 3111–3119, Red Hook, NY, USA, 2013. Curran Associates
  Inc.

\bibitem{DBLP:conf/nips/MikolovSCCD13}
Tom{\'{a}}s Mikolov, Ilya Sutskever, Kai Chen, Gregory~S. Corrado, and Jeffrey
  Dean.
\newblock Distributed representations of words and phrases and their
  compositionality.
\newblock In Christopher J.~C. Burges, L{\'{e}}on Bottou, Zoubin Ghahramani,
  and Kilian~Q. Weinberger, editors, {\em Advances in Neural Information
  Processing Systems 26: 27th Annual Conference on Neural Information
  Processing Systems 2013. Proceedings of a meeting held December 5-8, 2013,
  Lake Tahoe, Nevada, United States}, pages 3111--3119, 2013.

\bibitem{type4pydataset}
Amir~M. Mir, Evaldas Latoskinas, and Georgios Gousios.
\newblock Manytypes4py: {A} benchmark python dataset for machine learning-based
  type inference.
\newblock {\em CoRR}, abs/2104.04706, 2021.

\bibitem{mir2021type4py}
Amir~M Mir, Evaldas Latoskinas, Sebastian Proksch, and Georgios Gousios.
\newblock Type4py: Deep similarity learning-based type inference for python.
\newblock {\em arXiv preprint arXiv:2101.04470}, 2021.

\bibitem{Yusuke19dynamictypeinfer}
Yusuke Miyazaki, Taro Sekiyama, and Atsushi Igarashi.
\newblock Dynamic type inference for gradual hindley-milner typing.
\newblock {\em Proc. {ACM} Program. Lang.}, 3({POPL}):18:1--18:29, 2019.

\bibitem{mypy}
Mypy.
\newblock https://github.com/python/mypy/.

\bibitem{Zvonimir21dataflowtypeinfer}
Zvonimir Pavlinovic, Yusen Su, and Thomas Wies.
\newblock Data flow refinement type inference.
\newblock {\em Proc. {ACM} Program. Lang.}, 5({POPL}):1--31, 2021.

\bibitem{typewriter}
Michael Pradel, Georgios Gousios, Jason Liu, and Satish Chandra.
\newblock {\em TypeWriter: Neural Type Prediction with Search-Based
  Validation}, page 209–220.
\newblock Association for Computing Machinery, New York, NY, USA, 2020.

\bibitem{pyre}
Pyre check.
\newblock https://pyre-check.org/.

\bibitem{pyright}
Pyright.
\newblock https://github.com/microsoft/pyright.

\bibitem{pysonar}
Pysonar2.
\newblock https://github.com/yinwang0/pysonar2.

\bibitem{pyast}
Python.
\newblock The python ast module, 2021.
\newblock \url{https://github.com/python/cpython/blob/3.9/Lib/ast.py}.

\bibitem{typeshed}
Python.
\newblock The typeshed project, 2021.
\newblock https://github.com/python/typeshed.

\bibitem{pytype}
Pytype.
\newblock https://github.com/google/pytype.

\bibitem{raunak-etal-2020-long}
Vikas Raunak, Siddharth Dalmia, Vivek Gupta, and Florian Metze.
\newblock On long-tailed phenomena in neural machine translation.
\newblock In {\em Findings of the Association for Computational Linguistics:
  EMNLP 2020}, pages 3088--3095, Online, November 2020. Association for
  Computational Linguistics.

\bibitem{Ray17study}
Baishakhi Ray, Daryl Posnett, Premkumar Devanbu, and Vladimir Filkov.
\newblock A large-scale study of programming languages and code quality in
  github.
\newblock {\em Commun. ACM}, 60(10):91–100, September 2017.

\bibitem{jsnice}
Veselin Raychev, Martin Vechev, and Andreas Krause.
\newblock Predicting program properties from "big code".
\newblock {\em SIGPLAN Not.}, 50(1):111–124, January 2015.

\bibitem{rubychecker13}
Brianna~M. Ren, John Toman, T.~Stephen Strickland, and Jeffrey~S. Foster.
\newblock The ruby type checker.
\newblock In {\em Proceedings of the 28th Annual ACM Symposium on Applied
  Computing}, SAC '13, page 1565–1572, New York, NY, USA, 2013. Association
  for Computing Machinery.

\bibitem{ren2020balanced}
Jiawei Ren, Cunjun Yu, Shunan Sheng, Xiao Ma, Haiyu Zhao, Shuai Yi, and
  Hongsheng Li.
\newblock Balanced meta-softmax for long-tailed visual recognition, 2020.

\bibitem{jsdynamic}
Gregor Richards, Sylvain Lebresne, Brian Burg, and Jan Vitek.
\newblock An analysis of the dynamic behavior of javascript programs.
\newblock PLDI '10, page 1–12, New York, NY, USA, 2010. Association for
  Computing Machinery.

\bibitem{pycg}
Vitalis Salis, Thodoris Sotiropoulos, Panos Louridas, Diomidis Spinellis, and
  Dimitris Mitropoulos.
\newblock Pycg: Practical call graph generation in python.
\newblock In {\em 43rd {IEEE/ACM} International Conference on Software
  Engineering, {ICSE} 2021, Madrid, Spain, 22-30 May 2021}, pages 1646--1657.
  {IEEE}, 2021.

\bibitem{BPE16}
Rico Sennrich, Barry Haddow, and Alexandra Birch.
\newblock Neural machine translation of rare words with subword units.
\newblock In {\em Proceedings of the 54th Annual Meeting of the Association for
  Computational Linguistics (Volume 1: Long Papers)}, pages 1715--1725, Berlin,
  Germany, August 2016. Association for Computational Linguistics.

\bibitem{cauldron}
sernst.
\newblock The return value at line 35., 2021.
\newblock
  https://github.com/sernst/cauldron/blob/master/cauldron/steptest/functional.py.

\bibitem{growser}
tomdean.
\newblock The return value at line 56., 2021.
\newblock
  https://github.com/tomdean/growser/blob/master/growser/handlers/rankings.py.

\bibitem{pep484}
Guido van Rossum, Jukka Lehtosalo, and Łukasz Langa.
\newblock {PEP} 484 -- {Type Hints}, 2014.
\newblock https://www.python.org/dev/peps/pep-0484/.

\bibitem{wei2020lambdanet}
Jiayi Wei, Maruth Goyal, Greg Durrett, and Isil Dillig.
\newblock Lambdanet: Probabilistic type inference using graph neural networks.
\newblock {\em CoRR}, abs/2005.02161, 2020.

\bibitem{metadata-check}
wtsi hgi.
\newblock The return value at line 151., 2021.
\newblock
  https://github.com/wtsi-hgi/metadata-check/blob/master/mcheck/metadata/seqscape\_metadata/seqscape\_metadata.py.

\bibitem{pbinfer16}
Zhaogui Xu, Xiangyu Zhang, Lin Chen, Kexin Pei, and Baowen Xu.
\newblock Python probabilistic type inference with natural language support.
\newblock In {\em Proceedings of the 2016 24th ACM SIGSOFT International
  Symposium on Foundations of Software Engineering}, FSE 2016, page 607–618,
  New York, NY, USA, 2016. Association for Computing Machinery.

\bibitem{zhang-etal-2019-long}
Ningyu Zhang, Shumin Deng, Zhanlin Sun, Guanying Wang, Xi~Chen, Wei Zhang, and
  Huajun Chen.
\newblock Long-tail relation extraction via knowledge graph embeddings and
  graph convolution networks.
\newblock In {\em Proceedings of the 2019 Conference of the North {A}merican
  Chapter of the Association for Computational Linguistics: Human Language
  Technologies, Volume 1 (Long and Short Papers)}, pages 3016--3025,
  Minneapolis, Minnesota, June 2019. Association for Computational Linguistics.

\bibitem{pep585}
Łukasz Langa.
\newblock {PEP} 589 -- type hinting generics in standard collections, March
  2019.
\newblock https://www.python.org/dev/peps/pep-0585/.

\bibitem{gensim}
Radim Řehůřek and Petr Sojka.
\newblock Software framework for topic modelling with large corpora.
\newblock pages 45--50, 05 2010.

\end{thebibliography}

\end{document}